\documentclass[reprint, aps, prd, superscriptaddress]{revtex4-2}

\usepackage{graphicx}
\usepackage{dcolumn}
\usepackage{bm}
\usepackage{subcaption}
\usepackage{makecell}
\usepackage{amsmath}
\usepackage{amssymb}
\usepackage{tikz}
\usepackage{placeins}
\usepackage{aas_macros}

\begin{document}

\preprint{APS/123-QED}

\title{Systematic Study of Forward and Reverse Shock Afterglow Emission from Two-Component Jets}

\author{Olzhas Mukazhanov}
\email{olzhmuk@gmail.com}
\affiliation{Department of Physics, Nazarbayev University, 53 Kabanbay Batyr Ave, Astana 010000, Kazakhstan}
\affiliation{Energetic Cosmos Laboratory, Nazarbayev University, 53 Kabanbay Batyr Ave, Astana 010000, Kazakhstan}

\author{Ernazar Abdikamalov}
\affiliation{Department of Physics, Nazarbayev University, 53 Kabanbay Batyr Ave, Astana 010000, Kazakhstan}
\affiliation{Energetic Cosmos Laboratory, Nazarbayev University, 53 Kabanbay Batyr Ave, Astana 010000, Kazakhstan}

\author{Paz Beniamini}
\affiliation{Department of Natural Sciences, The Open University of Israel, PO Box 808, Ra’anana 4353701, Israel}
\affiliation{Astrophysics Research Center of the Open university (ARCO), The Open University of Israel, PO Box 808, Ra’anana 4353701, Israel}
\affiliation{Department of Physics, The George Washington University, 725 21st Street NW, Washington, DC 20052, USA}

\date{\today}

\begin{abstract}
Two-component jets are frequently invoked to explain complex features in gamma-ray burst (GRB) afterglows, such as late-time rebrightening and chromatic breaks. While many studies fit these models to individual events, a systematic exploration mapping the broader parameter space, particularly the reverse shock contribution, is currently lacking. To address this, we present a comprehensive systematic analysis of two-component jet signatures using numerical modeling with the VegasAfterglow code. Our modeling shows that observable rebrightenings in the forward shock require the wing to carry substantially more energy, while for the reverse shock the energies can be comparable. Because the two components can occupy different spectral regimes, spectral breaks may arise when the wing emission overtakes the core. When the wing's initial velocity is high, relativistic beaming can render its emission invisible to the on-axis observer. As the flow decelerates, the resulting debeaming produces a steeper rise in the observed emission, reaching temporal slopes as steep as about $4.5$ and peaking shortly after the core jet break. In this case, the wing masks the core's break, leaving only a single late-time break. Slower wings that are not initially beamed away do not obscure the core, allowing the observer to see two distinct jet breaks. At late times, the decaying post-jet-break slopes are unaffected and limited to temporal slopes of about $-p$. Additionally, the forward shock dominates the emission across most of the parameter space, while the reverse shock contributes noticeably only under conditions of high magnetization and long engine durations.
\end{abstract}

\maketitle

\section{Introduction}
\label{sec:intro}

Gamma-ray bursts (GRBs) are the most luminous electromagnetic explosions in the Universe. The standard framework for describing GRB phenomenology is the relativistic fireball model \citep{Rees92Relativistic, Piran99Gamma}. In this picture, a central engine launches an ultra-relativistic, collimated outflow that dissipates energy to produce the prompt $\gamma$-ray emission \citep{Rees94Unsteady, Kobayashi97Can, Piran04Review}. Subsequently, the ejecta decelerates as it sweeps up the circumburst medium, driving a relativistic forward shock (FS) into the ambient medium and a reverse shock (RS) back into the ejecta \citep{Sari95AHydrodynamic, Meszaros93Relativistic, Kobayashi99Hydrodynamics}. The synchrotron and synchrotron self-Compton radiation emitted by electrons accelerated in these shocks produces the multi-wavelength afterglow emission \cite{Meszaros97Optical, Sari98Spectra, Sari99Predictions, Zhang03Gamma, Zhang06Physical}.

While the standard fireball model successfully explains the overall temporal and spectral properties of many GRB afterglows, the era of detailed monitoring ushered in by Swift and Fermi has revealed a much richer phenomenology. Enabled by fast-slewing ground-based robotic telescopes \citep{Lipunov10Master, Burd05Pi, Vestrand06Energy, Klotz09Early, Komesh2023Evolution, Abdullayev25Early, Akerlof99Observation} and new missions such as SVOM \citep{Wei16SVOM} and the Einstein Probe \citep{Yuan22EP}, these observations show complex light-curve features that deviate from simple power-law decays \citep{Nousek06Evidence, Wang15How}. These modern facilities, capable of rapid follow-up and wide-field monitoring, have been instrumental in characterizing steep decay phases, shallow decay plateaus, chromatic breaks, and flares \citep{Wang15How, Yi22Statistical, Busmann25Curious, Akl26Multi}. One notable class of phenomena is late-time rebrightenings, or “bumps,” in the optical and radio bands, observed hours to weeks after the trigger \citep{Liang2013, Nardini2014}. Such bumps challenge the standard single-component jet model, often requiring modifications such as energy injection \citep{Sari00Impulsive, Zhang02GRB, Panaitescu06Analysis, Laskar15Energy, AnguloValdez26Evidence}, density variations in the circumburst medium \cite{Dai02Hydrodynamics, Ren25Stellar, Nakar07Smooth}, variations in microphysical parameters \citep{Filgas11GRB091127, Panaitescu06Evidence, Misra21Low}, jet structure \citep{Lamb17Electromagnetic,Beniamini22structure, OConnor23Structured, Gill23GRB, Salafia22Structure}, or off-axis viewing \citep{Meszaros98Viewing, Granot02Off, Rossi02Afterglow, Panaitescu03Effect,Beniamini20MAfterglow,Beniamini20Plateau,Birenbaum24Afterglow, Abdikamalov25Reverse, Wang25Forward}.

One of the simplest explanations for these complex morphologies is the two-component jet model \cite{RamirezRuiz02Beam}. Theoretical considerations indicate that GRB outflows are unlikely to be uniform "tophat" structures; instead, they naturally develop transverse stratification consisting of a fast core and a slower wing \cite{Salafia22Structure}.

This structure is supported by numerical simulations. Magnetohydrodynamic models of the central engine suggest that magnetic stresses can launch a relativistic inner jet alongside a baryon-loaded outer wind \cite{McKinney06General, Komissarov09Magnetic, Desai26Relativistic}. Moreover, as the jet propagates through the dense stellar envelope (in long GRBs) or merger ejecta (in short GRBs), the interaction with the surrounding medium creates a high-pressure cocoon \cite{Morsony07Temporal, Bromberg11Propagation, Gottlieb25Landscape}. This process sheathes the ultra-relativistic core in a wider, slower-moving layer of shocked material \cite{Duffell15From, Tchekhovskoy08Simulations}.

While these physical mechanisms often yield a continuous jet structure with energy and Lorentz factor decreasing smoothly toward the wings, the two-component model serves as a simple approximation \cite{Zhang04Quasi, Xie12two}. In this scenario, the initial afterglow is dominated by emission from the highly beamed core. As the core decelerates, the emission from the wider sheath becomes visible, producing the characteristic flattening or distinct rebrightening observed in late-time light curves \cite{Granot05Afterglow, Peng05Two}.

There is extensive observational evidence for two-component jet structure. Early indications arose from studies of GRB 970508, where deviations from simple spherical blast-wave evolution were identified \citep{Pedersen98Evidence}. Subsequently, detailed multi-wavelength observations provided more direct confirmation. Similarly, energetics and beaming analyses of GRB 991216 \citep{Frail00Enigmatic} highlighted the possibility that different angular regions of the jet could contribute differently to the prompt and afterglow emission. A decisive test came from the radio calorimetry of GRB 030329, where \citet{Berger03Common} demonstrated that the data favored a two-component explosion, with the wider sheath dominating the radio and optical emission at times $ \gtrsim 1.5$ days \citep{Sheth03Millimeter}.

In recent years, the model has proved useful for explaining complex spectral and temporal features. For instance, it has been used to explain chromatic breaks in GRB 191221B \citep{Chen24GRB191221B} and inconsistent jet break times in GRB 230815A \citep{Leung2025}. It also accommodates high-energy anomalies; \citet{Sato25Two} showed that the $\gtrsim 100$ GeV afterglows of GRBs 201216C and 221009A are consistent with a two-component scenario \citep{Sato23Synchrotron, Sato23Two}. Further evidence for angular structure comes from GRB 221009A (the BOAT), where modeling favor a structured jet with an energetic and narrow core \citep{Laskar2023BOAT, Zhang2024BOAT} (although see \cite{OConnor23Structured,Gill23GRB} for an alternate interpretation). While the core carries a large isotropic-equivalent energy, its small opening angle implies that the jet wings may dominate the true (beaming-corrected) energy budget. In this picture, the ultra-relativistic core powers the prompt emission, whereas the wider and less relativistic sheath contributes substantially to the afterglow. Such angular energy stratification can lead to apparent discrepancies in inferred GRB radiative efficiencies when prompt and late-time afterglow energetics are compared under single-component assumptions. Beyond light-curve modeling, \citet{Wu05Gamma} explored the distinct polarization signatures expected from such structured jets. Consequently, two-component models are now frequently invoked to explain a wide range of afterglow behaviors, from the rebrightening of XRF 030723 \citep{Huang04Rebrightening} to the long-lived emission of GRB 111209A \citep{Kann2018, Filgas2011}, and numerous other cases \citep{Liang04Peak, Starling05Spectroscopy, Granot2006, Oates07two, Jin07Two, Nardini2011, Cucchiara15Happy, Chen20Two, Kangas21Late, Zhu24two, Tian26Short}.

Modeling these multi-component structures requires a formalism capable of handling non-uniform jet profiles. \citet{Granot05Afterglow} developed a semi-analytical formalism capable of calculating the flux from various jet geometries, including ring-shaped, fan-shaped, and two-component structures. This framework was subsequently applied by \citet{Peng05Two} to perform a detailed investigation of the forward shock emission in the two-component scenario. In recent years, the modeling of GRB afterglows has been advanced by numerical codes capable of treating arbitrary lateral structures and off-axis viewing angles \citep[e.g.,][]{vanEerten10Off, Gill18Afterglow, Ryan20Aafterglowpy, Sarin24Redback, Wang26VegasAfterglow}. To date, however, these studies have mainly focused on the forward shock emission, with the reverse shock receiving comparatively less attention in the context of spine-sheath geometries \citep{Lamb19Reverse, Pang24Reverse}.

Since the sheath carries significant energy but a lower Lorentz factor, its interaction with the circumburst medium is expected to produce emission that peaks at later times compared to the core. Furthermore, despite the popularity of the two-component model in fitting individual anomalous bursts, the parameter space has not been explored systematically. It remains unclear exactly where in the parameter space (e.g., specific ratios of core-to-sheath energy or opening angles) the interplay of these components results in a distinct rebrightening versus a mere broadening of the jet break.
  
In this work, we perform a systematic investigation of the afterglow emission from two-component jets, incorporating both the forward-shock and reverse-shock contributions. We utilize the numerical code \texttt{VegasAfterglow} \citep{Wang26VegasAfterglow} to simulate the radio, optical, and X-ray emission across a broad region of the parameter space.  Specifically, we vary key parameters—including the isotropic equivalent energy, initial Lorentz factor, and jet opening angle—for both the core and the wing components. To illuminate the underlying physics driving the numerical results, we complement our simulations with analytical scalings. Our primary goal is to identify the specific physical conditions under which distinct signatures, such as rebrightenings or flattening, become observable against the standard decay. This analysis serves as a guide for constraining the role of lateral jet structure in shaping the observed diversity of GRB afterglows.

The paper is organized as follows. In Section~\ref{sec:method}, we describe the numerical framework and the physical model used for the two-component jet simulations. In Section~\ref{sec:results}, we present the resulting light curves and analyze the dependency of the emission features on the core and wing parameters. Finally, in Section~\ref{sec:conclusion}, we summarize our findings and discuss their implications for interpreting future multi-wavelength observations.

\section{Method}
\label{sec:method}

\begin{figure*}[ht]
    \centering
    \includegraphics[width=\textwidth]{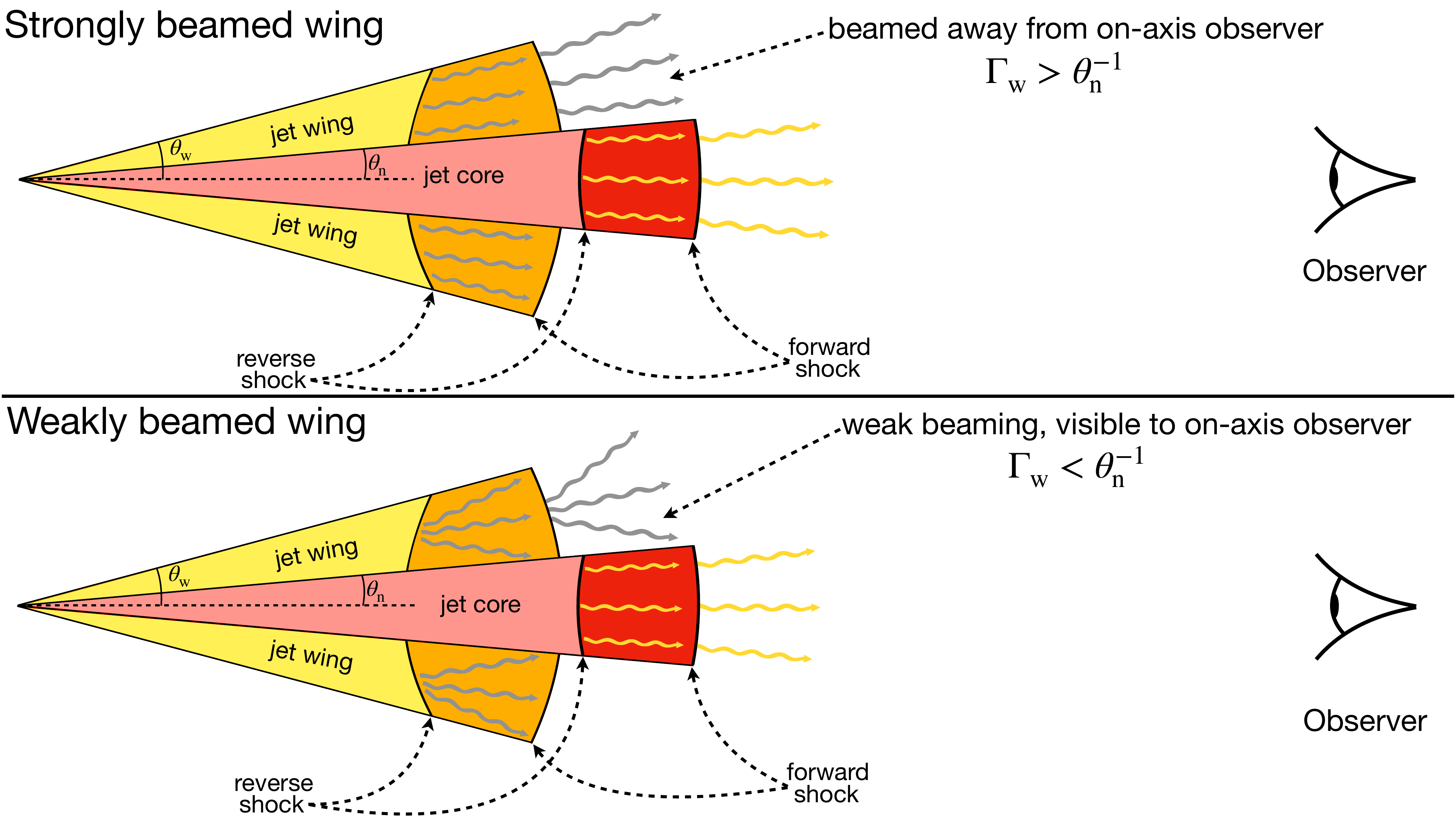}
    \caption{Schematic illustration of a two-component jet structure and the associated forward–reverse shock pairs formed in both the jet core and the wing. The top panel depicts the case of a fast wing, in which the emission is relativistically beamed away from an on-axis observer. The bottom panel shows the case of a slower wing, where the wing emission lies within the observer’s beaming cone and is therefore detectable. The wavy arrows indicate the radiation originating from the forward- and reverse-shocked regions.}
    \label{fig:scheme}
\end{figure*}

\begin{table}[t]
\centering
\caption{Default model parameters for ISM and wind medium.}
\begin{tabular}{lc}
\hline\hline
Parameter & Default value \\ 
\hline
\multicolumn{2}{c}{\textbf{Microphysical parameters}} \\
$\varepsilon_e$ & $10^{-1}$ \\
$\varepsilon_B$ & $10^{-3}$ \\
$p$ & 2.3 \\
\hline
\multicolumn{2}{c}{\textbf{Observer parameters}} \\
Distance & $10^{28}$ cm \\
$\theta_{\rm obs}$ & 0 \\
\hline
\multicolumn{2}{c}{\textbf{Medium}} \\
$n_{\rm ISM}$ & 1 cm$^{-3}$ \\
$A_*$ & 0.1 \\
\hline
\multicolumn{2}{c}{\textbf{Jet parameters}} \\
$\Gamma_{\rm 0,(n;w)}$ & 150; 100 \\
$\theta_{\rm (n,w)}$ & 0.05; 0.2 \\
$E_\mathrm{ISO,(n;w)}$ & $10^{53}$ erg \\
$T_{\rm eng}$ & 10 s \\
\hline\hline
\end{tabular}
\label{tab:const_params}
\end{table}

We use a combination of analytical and numerical approach. For the latter, we use the \texttt{VegasAfterglow} code, which is designed to compute broadband afterglow emission from GRB  jets \citep{Wang26VegasAfterglow}. It solves the synchrotron radiation from relativistic ejecta interacting with the circumburst medium, accounting for both forward and reverse shock contributions. The code allows flexible specification of jet geometry, including tophat, power-law, Gaussian, and user-defined structured jets, as well as various external density profiles such as uniform interstellar medium or stellar wind for for arbitrary observer viewing angles. See \citep{Wang26VegasAfterglow} for the code details. 

We study emission from both forward and reverse shocks. We explore a wide range of jet parameters for both the jet core and wing, including energy, Lorentz factor, and opening angle. We also consider both constant-density ISM and wind-like external media (and transitions between the two), as well as different ejecta durations. As a result, we cover a wide range of scenarios, covering thin and thick shells cases. Together, these choices already introduce a large number of degrees of freedom. To keep the parameter space manageable, we restrict our analysis to on-axis observers and assume that the microphysical parameters $\varepsilon_e$, $\varepsilon_B$, and $p$ are identical in the jet core and wing. Here, $\varepsilon_e$ and $\varepsilon_B$ denote the fractions of the post-shock internal energy density imparted to accelerated electrons and magnetic fields, respectively, and $p$ is the power-law index of the electron energy distribution. We further assume that the afterglow emission is dominated by synchrotron radiation from shock-accelerated electrons, neglecting inverse Compton scattering (both its emission and its effect on electron cooling) and other radiative processes. Finally, we neglect lateral expansion of the jet and the interaction between the components of the jet. These limitations will be addressed in future works.

We approach the parameter study by selecting a default model and varying individual parameters independently. To ensure the robustness of our results, we tested several default configurations, including cases with wind and ISM mediums as well as transitions between the two. We find qualitatively similar systematics. The default parameter choice adopted for presenting our results in this paper is listed in Table~\ref{tab:const_params} for both ISM and wind mediums. We analyze the resulting light curves in three different bands: radio at 5 GHz, optical R-band at 650 nm, and X-ray and 5 keV. While the explored parameter values that span a wide numerical range, our primary objective is not to determine the physical likelihood of any specific parameter set, but rather to systematically quantify the impact of each variable on the resulting emission.   

The emission from a two-component jet is a superposition of the emissions from its components. See Fig.~\ref{fig:scheme} for schematic illustration of a two-component jet. Different components may peak at different times. Following the convention in the literature, we classify the peaks into three types: spectral (when a particular band crosses the critical frequencies), dynamic (when the shock finishes crossing the shell), and geometric (caused by debeaming as the off-axis wing component decelerates) \citep[e.g.,][]{Sari98Spectra, Sari99Predictions, Kobayashi00Light, Kobayashi03Early, Gao15Morphological}. Throughout this work, the subscripts "n" and "w" denote quantities associated with the narrow (core) and wide (wing) components, respectively.

The pre-peak slopes for each component are derived from analytical scaling formulas provided in Appendix~\ref{sec:formulas}. For the wing, a geometric correction is required. Although the observer lies on the jet symmetry axis, all wing material is viewed at an angle because the wing begins at $\theta_{\rm n}$ (cf. Fig.~\ref{fig:scheme}). The dominant contribution to the pre-peak and peak emission arises from the portion of the wing closest to the line of sight, i.e. at its inner edge $\theta_{\rm n}$. Emission from larger angles is both Doppler-suppressed and geometrically delayed to a stronger extent, so it contributes primarily at later times. 

We therefore compare the Doppler factor of the wing's inner edge as seen by the on-axis observer (for whom the edge is viewed off-axis) to the Doppler factor that would be measured by a hypothetical observer aligned with the edge's velocity. We define
\begin{equation}
    a(\Gamma_{\rm w}) \equiv \frac{\mathcal{D}(\theta_{\rm n})}{\mathcal{D}(0)} \approx \frac{1}{1+\Gamma_{\rm w}^2\theta_{\rm n}^2},
\end{equation}
where $\mathcal{D}(\theta_{\rm n})$ corresponds to the global on-axis observer, while $\mathcal{D}(0)$ corresponds to the observer aligned with the wing's inner edge.

For brevity in comparisons between the wing's and core's quantities, we define a ratio $\mathcal{R}(\mathcal{Q})$ of quantity $\mathcal{Q}$ for the wing and jet core, 
\begin{equation}
\mathcal{R}(\mathcal{Q}) \equiv \frac{\mathcal{Q}_{\rm w}}{\mathcal{Q}_{\rm n}},
\end{equation}
which will be used in our subsequent analysis.

\section{Results}
\label{sec:results}

\subsection{Qualitative Picture}
\label{sec:qualitative}

The qualitative picture of emission from two-component jets is well understood \citep{Granot2006, Peng05Two}. Nevertheless, for completeness, we briefly highlight the key aspects below.

Consider a cold ejecta fired into circumburst gas. When this jet slams into the gas, it creates two shocks: outward-moving FS and backward-moving RS. This creates four regions: (1) unshocked circumburst gas ahead of FS, (2) the shocked circumburst gas, (3) shocked ejecta: the jet's material, heated and compressed by the RS, and (4) unshocked ejecta: the innermost jet material that has not yet collided with the RS. Sandwiched between the two shocks, regions 2 and 3 push against each other \citep[e.g.,][]{Sari95AHydrodynamic, Kobayashi99Hydrodynamics}. 

The evolution of the system depends on whether the RS has time to become relativistic before crossing the ejecta, creating two scenarios: thin or thick shell cases \citep[e.g.,][]{Sari95AHydrodynamic}. In a thick shell, the RS has time to become relativistic. It brakes the ejecta, rapidly dropping its speed. The blast wave speed now depends entirely on the jet total power, not its initial speed. The RS extracts most of the jet's kinetic energy, dumping it into the shocked gas. After crossing the shell, the FS smoothly decelerates following standard Blandford-McKee solution \citep{BM76}.

In a thin shell, the RS crosses the ejecta before becoming relativistic. Initially, the RS has little impact on the overall dynamics of the jet, cruising at its starting speed. It only decelerates after scooping up sufficient amount of circumburst gas, equal to its own mass divided by its initial Lorentz factor. From there, the FS begins its Blandford-McKee slowdown, but the jet material remains relatively cool because RS never became relativistic \citep[e.g.,][]{Kobayashi99Hydrodynamics}. 

\begin{figure*}[ht]
    \centering
    \begin{subfigure}[t]{0.45\textwidth}
        \centering
        \includegraphics[width=\linewidth]{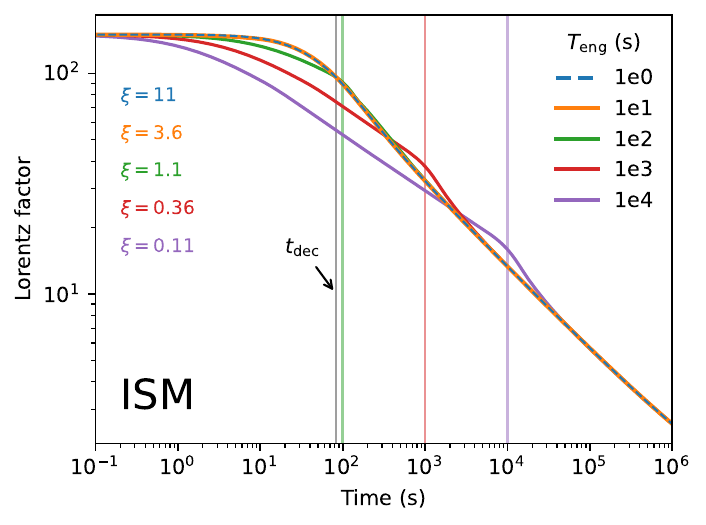}
        \label{fig:Gamma_dur_ISM}
    \end{subfigure}
    \begin{subfigure}[t]{0.45\textwidth}
        \centering
        \includegraphics[width=\linewidth]{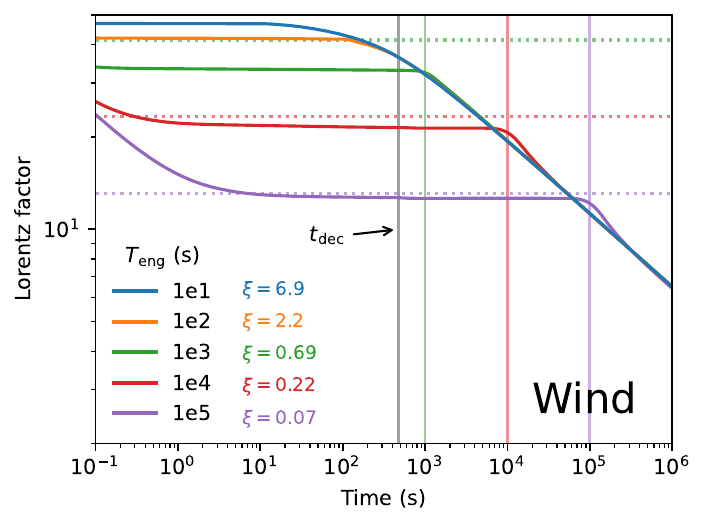}
        \label{fig:Gamma_dur_Wind}
    \end{subfigure}
    \caption{Evolution of Lorentz factor of the forward-shocked region for different engine durations. The initial Lorentz factor is set at $\Gamma_0=150$ and $\Gamma_0=50$ for the ISM and wind mediums, respectively. Shell thickness is determined by parameter $\xi$ with $\xi>1$ corresponding to `thin shell' and vice versa. The colored vertical lines on both panels depict $T_{\rm eng}$. The colored horizontal dotted lines on the right panel show the critical Lorentz factor $\Gamma_\mathrm{c}$ for thick-shell cases. 
    } 
    \label{fig:Gamma_dur}
\end{figure*}

Figure~\ref{fig:Gamma_dur} shows the time evolution of the Lorentz factor of the forward-shocked region for different engine durations $T_\mathrm{eng}$ and therefore effective shell thicknesses for ISM (left panel) and wind (right panel) media. In the thin-shell regime ($T_{\rm eng} < t_{\rm dec}$ or $\xi > 1$), the blastwave initially moves at a constant speed until the reverse shock crosses the ejecta, after which the ejecta transitions to a self-similar deceleration \citep{Kobayashi99Hydrodynamics, GovreenSegal24}.

However, the thick-shell regime ($T_{\rm eng} > t_{\rm dec}$ or $\xi < 1$) yields a different result for the ISM and the wind. In the ISM, the reverse shock becomes relativistic early and slows down the blastwave during the crossing phase, after which the Lorentz factor decelerates self-similarly as with the thin shell \citep{Zhang22Semi}. In the wind, the Lorentz factor quickly slows down to a critical value $\Gamma_{\rm c} < \Gamma_0$ and stays constant until $T_{\rm eng}$, after which it decelerates as in the thin-shell regime \citep{Zou05Early}.

In a two-component jet, both the fast narrow core and the slower, wider wing undergo this shock evolution. Because of their differing initial Lorentz factors and energy densities, the two components can operate in different shell regimes, resulting in a complex superposition of their forward and reverse shock emissions. The core emission behaves as a standard on-axis tophat jet, while the wing is treated like a ring around symmetry axis \cite{Granot05Afterglow}. 

In general, the observed emission depends on the frequency band, which may fall into different spectral regimes. For concreteness, consider here optical emission, which typically lies above both $\nu_\mathrm{m}$ and $\nu_\mathrm{a}$ over most of the relevant parameter space \citep{Gao15Morphological}. In this spectral regime, the observed peak can be either dynamical (associated with shock crossing) or geometric (associated with debeaming). For the core, the peak occurs at the deceleration time $t_{\rm dec}$ in the thin-shell case and at $T_{\rm eng}$ in the thick-shell regime. For the wing, the situation depends on its Lorentz factor. Because the wing's inner edge is located off the symmetry axis, its emission may initially be Doppler-suppressed \citep{Granot2006}. If its Lorentz factor remains larger than $1/\theta_{\rm n}$ at shock crossing, the dynamical forward-shock peak is hidden and the observed flux maximum occurs later at time $t_{\rm beam, w}$, when the Lorentz factor decreases to $\Gamma_{\rm w}\sim1/\theta_{\rm n}$ (cf. top panel of Fig.~\ref{fig:scheme}). Conversely, if the wing is slow enough at the shock-crossing time ($\Gamma_{\rm w} \lesssim 1/\theta_{\rm n}$), its dynamical FS peak is directly visible (cf. bottom panel of Fig.~\ref{fig:scheme}). 

The wing's reverse shock behaves differently. Once the shock completes its passage through the ejecta, the emission declines rapidly \citep{Kobayashi00Light}. This steep post-crossing decay suppresses the formation of a later geometric debeaming-driven peak. Consequently, unlike in the forward shock, the wing's RS dynamical peak is never obscured by the debeaming effect. The emission peaks at the shock-crossing time in both thin- and thick-shell regimes. In the thin-shell case, the wing flux rises with the same temporal slope as the core, because the Lorentz factor stays constant before $t_{\rm dec}$ in the thin-shell regime. For thick shells, since the blastwave decelerates during the crossing phase, the flux slopes can be geometrically enhanced in the observer's frame.

In the following, we discuss in detail the radio, optical, and X-ray emission from forward and reverse shocks in both thin- and thick-shell regimes and in wind and ISM environments, including the transitions between them.

\subsection{Forward Shock in ISM case}
\label{sec:forward}

\begin{figure*}[p]
    \centering
    \includegraphics[width=\textwidth]{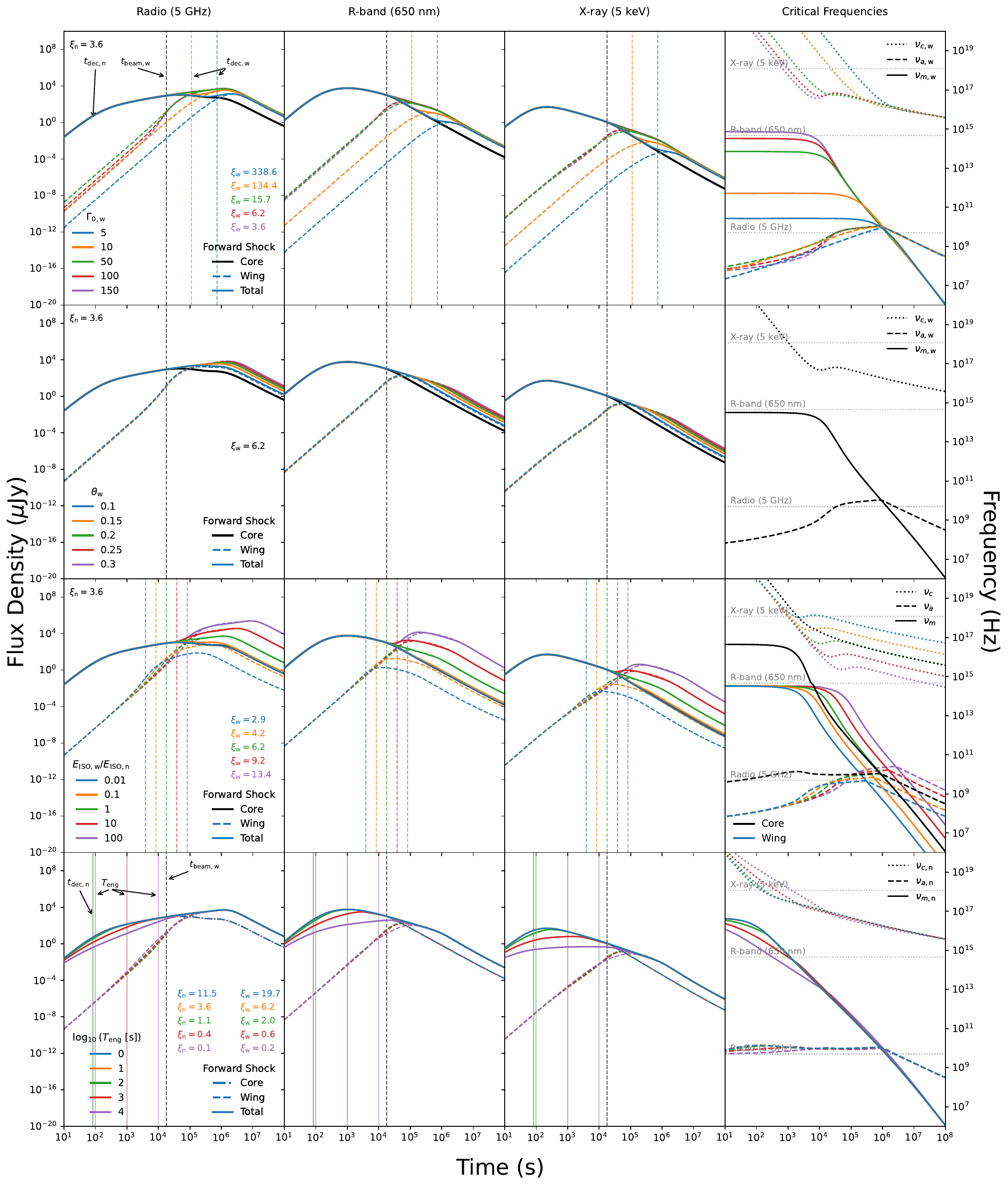}
    \caption{Multi-band light curves of forward shock for different values of the varied parameters for the ISM. Each row corresponds to a distinct parameter, while columns correspond to different frequency bands: radio (5 GHz), R-band (650 nm), and X-ray (5 keV). The solid vertical lines in the last row depict max$(t_{\rm dec}, T_{\rm eng})$.}
    \label{fig:forward}
\end{figure*}

Whether the wing emission is visible from the beginning depends on its initial Lorentz factor. If $\Gamma_{0,\rm w} > 1 / \theta_{\rm n}$, the emission is initially beamed away from the on-axis observer \citep{Granot02Off}. For our adopted core opening angle $\theta_{\rm n}=0.05$, this corresponds to $\Gamma_{0,\rm w}>20$. In such cases, the wing emission enters the line of sight only after the blastwave decelerates below Lorentz factor of $20$ and the relativistic beaming cone gets wide enough to cover the on-axis observer ($\Gamma_{\rm w}\theta_{\rm n}<1$). When this happens, the peak of the flux corresponds to the geometric peak, $t_{\rm peak}=t_{\rm beam}$. On the other hand, if $\Gamma_{0,\rm w} < 1 / \theta_{\rm n}$, the wing flux is seen from the start, and its dynamical peak can be observed at $t_{\rm peak}=t_{\rm dec}$. The Doppler factor correction for a thin-shell wing can be written as
\begin{equation}
    a(\Gamma_{\rm w}) \Big|_{t_{\rm peak}} \approx
    \begin{cases}
        1/(1 + \Gamma_{0,\rm w}^2\theta_{\rm n}^2), & \Gamma_{0,\rm w} < 1/\theta_{\rm n} \\
        1/2, & \Gamma_{0,\rm w} > 1/\theta_{\rm n}
    \end{cases}.
\end{equation}
The ratio of the wing relative to the core's peak time is given by
\begin{equation}
\begin{split}
    \mathcal{R}(t_{\rm peak}) &= \mathcal{R}(E_{\rm ISO})^{1/3} \times \\
    &\begin{cases}
        \mathcal{R}(\Gamma_0)^{-8/3} (1 + \Gamma_{0,\rm w}^2\theta_{\rm n}^2), & \Gamma_{0,\rm w} < 1/\theta_{\rm n} \\
       2(\Gamma_{0,\rm n}\theta_{\rm n})^{8/3}, & \Gamma_{0,\rm w} > 1/\theta_{\rm n}
    \end{cases}.
\end{split}
\label{eq:FSthintime}
\end{equation}
Analogously, we can compare the peak fluxes:
\begin{equation}
    \mathcal{R}(F_{\nu,\rm peak}) = \mathcal{F}\mathcal{R}(E_{\rm ISO}) \times
    \begin{cases}
        a(\Gamma_{0,\rm w})^{\frac{p+5}{2}}\mathcal{R}(\Gamma_0)^{2(p-1)} \\
        a(1/\theta_{\rm n})^{\frac{p+5}{2}} (\Gamma_{0,\rm n}\theta_{\rm n})^{2(1-p)}
    \end{cases}, \label{eq:FSthinflux}
\end{equation}
where $\mathcal{F}$ is a factor that accounts for the wing's ring-shaped emitting area (see Eq.~\ref{eq:F}). From the equations above, we can see that for $\Gamma_{0,\rm w}>1/\theta_{\rm n}$ the wing's peak time and flux do not depend on its initial Lorentz factor. Therefore, we can expect the light curves for $\Gamma_{\rm 0, w}>20$ to be similar to each other.

In case of rebrightenings caused by the wing's rising flux, it is helpful to analyze the pre-peak slope. For the thin-shell case, it can be expressed as
\begin{align}
    \frac{d\log F_\nu^{\rm w}}{d\log t} \Bigg|_{t < t_{\rm peak}} &=
        \begin{cases}
            3 \\
            -\frac{3(p-1)}{4} + \frac{3(13-p)}{4}\frac{\Gamma_{\rm w}^2\theta_{\rm n}^2}{4 + \Gamma_{\rm w}^2\theta_{\rm n}^2}
        \end{cases} \nonumber \\
        &\approx
        \begin{cases}
            3 & (t < t_{\rm dec, w}) \\
            -1 + 8\frac{\Gamma_{\rm w}^2\theta_{\rm n}^2}{4 + \Gamma_{\rm w}^2\theta_{\rm n}^2} & (t_{\rm dec} < t < t_{\rm beam, w})
        \end{cases}, \label{eq:FSslopethin}
\end{align}
where, in the last approximate expression, we set $p=2.3$. In the limit when $\Gamma_{\rm w}\theta_{\rm n}\gg1$, the slope approaches $t^7$. 

Figure~\ref{fig:forward} shows FS light curves and critical frequencies for different values of four jet parameters: the initial Lorentz factor, the opening angle, the isotropic-equivalent energy, and the engine duration. Each parameter was varied one by one while others were fixed at their default values listed in Table~\ref{tab:const_params}. 

The first row of Fig.~\ref{fig:forward} shows FS light curves for wings with different initial Lorentz factors. We see that highly relativistic wings ($\Gamma_{0,\rm w} = 100,~150$) have a higher emission slope near the debeaming time $t_{\rm beam, w}$, raising the slope from $t^3$ up to $t^{4.5}$. In contrast, slower wings do not experience slope increase due to debeaming, producing the standard pre-peak slope of $t^3$ \cite{Gao13complete}. These results are consistent with Eq.~(\ref{eq:FSslopethin}). However, for our default parameter set the wing is not energetic enough to produce a clear rebrightening, and these rising slopes are obscured by the core emission (examples of more energetic wings producing a pronounced rebrightening are shown on the third row of Fig.~\ref{fig:forward}).

Sufficiently fast wings ($\Gamma_{0,\rm w}>20$) produce R and X-ray peaks that are set by the debeaming time rather than by the shock-crossing time. After the RS crosses the wing's shell, the blastwave Lorentz factor converges to the same deceleration solution, independent of its initial value \citep{BM76}. If debeaming occurs during this self-similar stage, the debeaming time $t_{\rm beam, w}$ is universal and independent of $\Gamma_{0,\rm w}$. The light curves of such wings are identical, which makes it harder to measure the wing's initial Lorentz factor from observations. This is consistent with the analytical predictions in Eqs.~(\ref{eq:FSthintime}) and~(\ref{eq:FSthinflux}), where the wing's peak flux and time are independent of $\Gamma_{0, \rm w}$ for initially beamed away cases.

In contrast, the wing flux for cases $\Gamma_{0,\rm w}=5,~10$ is visible from the start. These emissions reach their maximum at $t_{\rm dec, w}$ which is dependent on the initial Lorentz factor. As expected from Eqs.~(\ref{eq:FSthintime}) and~(\ref{eq:FSthinflux}), slower wings correspond to weaker and later peaks. In these cases, it is possible to probe $\Gamma_{0,\rm w}$ from the deceleration time $t_{\rm dec,w}$ given the energy and ISM density (see Eq.~\ref{eq:tdec} in the Appendix~\ref{sec:formulas}), similar to how it is commonly done in tophat jets \cite{Sari97Hydrodynamics, Sari99Predictions, Molinari07}. However, since the inner part of the ring is located at an angle to the observer, its arrival time will be delayed compared to the on-axis tophat case. This geometric correction should be taken into account (see Eqs.~\ref{eq:D} and~\ref{eq:Off} in the Appendix~\ref{sec:formulas}).

The second row of Fig.~\ref{fig:forward} shows the flux density for different values of the wing's opening angle $\theta_{\rm w}$. As expected, models with smaller $\theta_{\rm w}$ have earlier jet break times \citep{Rhoads99Dynamics, Frail01Beaming, Berger03Standard, Zhao20Statistical}. Besides this effect, we do not see any significant impact of $\theta_{\rm w}$. In particular, the peak time, flux, and slope are not affected by variations in $\theta_{\rm w}$. Analytical expressions also support that, as $\theta_{\rm w}$ does not appear in Eqs.~(\ref{eq:FSthintime}) and~(\ref{eq:FSthinflux}). This is expected since most of the observable emission comes from the inner regions of the wing.

The third row of Fig.~\ref{fig:forward} shows the flux density for different values of the wing energy. As expected, wings with larger energy are brighter \citep{Peng05Two}. As $E_{\rm ISO, w}$ increases, the peak time and flux shift to larger values in agreement with Eqs.~(\ref{eq:FSthintime}) and~(\ref{eq:FSthinflux}). Additionally, changes in the energy density bear no effect on the flux slopes, as expected from Eq.~(\ref{eq:FSslopethin}). However, the visibility of the wing's pre-peak flux can be obscured by emission from the core if it is bright enough.

When the wing isotropic-equivalent energy is smaller than that of the core, the wing's contribution to the light curve is difficult to distinguish. At late times, however, the relevant quantity is the total energy rather than $E_{\rm ISO}$. Because the wing has a larger opening angle, its true energy can exceed that of the core even if $E_{\rm ISO,w} < E_{\rm ISO,n}$. For small angles, the total energy scales roughly as $E_{\rm tot} \sim E_{\rm ISO}\theta^2$, so the ratio of wing to core energy is amplified by $(\theta_{\rm w}/\theta_{\rm c})^2$. For our default parameters, $\theta_{\rm n}=0.05$ and $\theta_{\rm w}=0.2$, this gives a factor of $\sim16$. If the energies are comparable or the wing is more energetic, the sheath emission can dominate the total flux at late times. In this case, the observed jet break may correspond to the wing component, occurring later than the core jet break, the signature of which is buried under emission from the wing \citep{Peng05Two}. Additionally, as the wing emission rises and overtakes the decaying core flux, it can produce a visible rebrightening, particularly prominent in the R and X-ray bands for $E_{\rm ISO, w} \gg E_{\rm ISO, n}$. In these bands, the core emission is already declining at the relevant epochs, so the increasing wing flux leads to a clear slope inversion. In contrast, in the radio band, because it is below $\nu_\mathrm{m}$, the core emission is typically still rising or near its peak \citep{Gao13complete}, so the wing's contribution mainly increases the overall flux without producing a pronounced a separate rebrightening feature in the light curve.

When the wing overtakes the core, a given observing band may probe different spectral regimes in the two components. If the microphysical parameters of the two components are identical, in the post-crossing phase where the evolution is self-similar, the characteristic frequencies of the two components differ only through their energy, with a more energetic wing having lower $\nu_c$ and higher $\nu_m$. As a result, an X-ray band can transition from $\nu < \nu_c$ (core-dominated) to $\nu > \nu_c$ (wing-dominated), producing a spectral slope change of $\Delta\beta=1/2$. At lower frequencies, such as in the near-infrared, the transition can occur from $\nu_m < \nu < \nu_c$ to $\nu_a < \nu < \nu_m$, leading to a stronger flattening, $\Delta\beta = -\frac{1}{3}-\frac{p-1}{2}$ ($\approx-0.98$ for $p=2.3$). In the radio band, the situation is more complex, as the observing frequency may lie near both $\nu_a$ and $\nu_m$, while these critical frequencies can themselves cross, leading to multiple spectral transitions over a short time interval.

The bottom panel of Fig.~\ref{fig:forward} shows the light curves for different central engine durations, ranging from 1 to $10^4$~s. The cases $T_{\rm eng}=1,~10,~100~\mathrm{s}$ correspond to thin shells, and longer durations produce thick shells. The thin-shell cases produce similar light curves, because the shock-crossing time is controlled by $t_{\rm dec}$ which is the same for all thin shells. In the thick-shell regime, the Lorentz factor slows down as $\Gamma\propto t^{-1/4}$ during the shock-crossing phase, and the dynamical peak is reached at $t_{\rm peak}=T_{\rm eng}>t_{\rm dec}$. In this case, the core peak flux scales as
\begin{equation}
    F^{\rm n}_{\nu,\rm peak} \propto E_{\rm ISO,n}^{\frac{p+3}{4}} T_{\rm eng}^{-\frac{3(p-1)}{4}}.
\end{equation}
In the bottom panel of Fig.~\ref{fig:forward}, we observe that the core R and X-ray peaks become progressively weaker as $T_{\rm eng}$ increases, in agreement with the scaling above. This reduction of the flux is a generic feature of thicker shells, as the injected energy is distributed over a longer engine activity time, leading to a weaker and more diluted emission during the shock-crossing phase. This is true for the radio band as well, even though it reaches its peak later after the shock crosses the shell.

For the same reason, the pre-peak slopes in the R and X-ray bands become shallower in thicker shells. The analytical thin-shell slope of $t^3$ quoted in Eq.~(\ref{eq:FSslopethin}) shifts to the thick-shell slope of $t^\frac{3-p}{2}=t^{0.35}$ for bands in the $\nu_\mathrm{m}<\nu<\nu_\mathrm{c}$ spectral regime. This is most clearly seen in the core's early X-ray light curves, before crossing $\nu_\mathrm{c}$. Nevertheless, this qualitative trend persists more generally: as $T_{\rm eng}$ increases, the flux slopes across all bands during the shock-crossing time become shallower.

In contrast, the wing's light curves in the thick-shell regime are similar to the thin cases. During the shock-crossing phase, the debeaming effect amplifies the thick-shell flux slope to
\begin{equation}
    \frac{d\log F_\nu^{\rm w}}{d\log t} \Bigg|_{t < t_{\times,\rm w}} = \frac{3-p}{2} + \frac{4\Gamma_{\rm w}^2\theta_{\rm n}^2}{2 + \Gamma_{\rm w}^2\theta_{\rm n}^2}, \label{eq:FSslopethick}
\end{equation}
approaching $4.3$ for $p=2.3$ in the limit $\Gamma_{\rm w}\theta_{\rm n}\gg1$. This is comparable to the thin-shell slope (see Eq.~\ref{eq:FSslopethin}). Even for the longest considered engine duration, $T_{\rm eng}=10^4~\mathrm{s}$, the debeaming time remains larger. Thus, at the moment the RS crosses the shell, the wing is still highly relativistic and beamed away from the observer, $\Gamma_{\rm w}(t_{\times,\rm w}) > 1/\theta_{\rm n}$. Therefore, the wing peak is purely geometric and occurs later during the self-similar phase, when $\Gamma_{\rm w}\theta_{\rm n}\sim1$. The corresponding peak time and flux are therefore identical to the thin-shell case and can be written as
\begin{align}
    t_{\rm peak,w} &\propto a(1/\theta_{\rm n})^{-1}E_{\rm ISO,w}^{1/3} \theta_{\rm n}^{8/3}, \\
    F^{\rm w}_{\nu,\rm peak} &\propto a(1/\theta_{\rm n})^{\frac{p+5}{2}}\mathcal{F}E_{\rm ISO,w} \theta_{\rm n}^{2(1-p)}.
\end{align}
Since $t_{\rm beam, w}>T_{\rm eng}$ in all cases considered here, the wing emission is insensitive to the engine duration. In contrast, the core emission is directly affected by $T_{\rm eng}$ in the thick-shell regime. However, for a different parameter set (not shown here) in which the wing enters the observer's line of sight during the shock-crossing phase ($t_{\rm beam} \lesssim T_{\rm eng}$), the wing emission would be expected to exhibit a similar dependence on $T_{\rm eng}$ as the core.

\subsection{Reverse Shock in ISM case}
\label{sec:reverse}

\begin{figure*}[p]
    \centering
    \includegraphics[width=\textwidth]{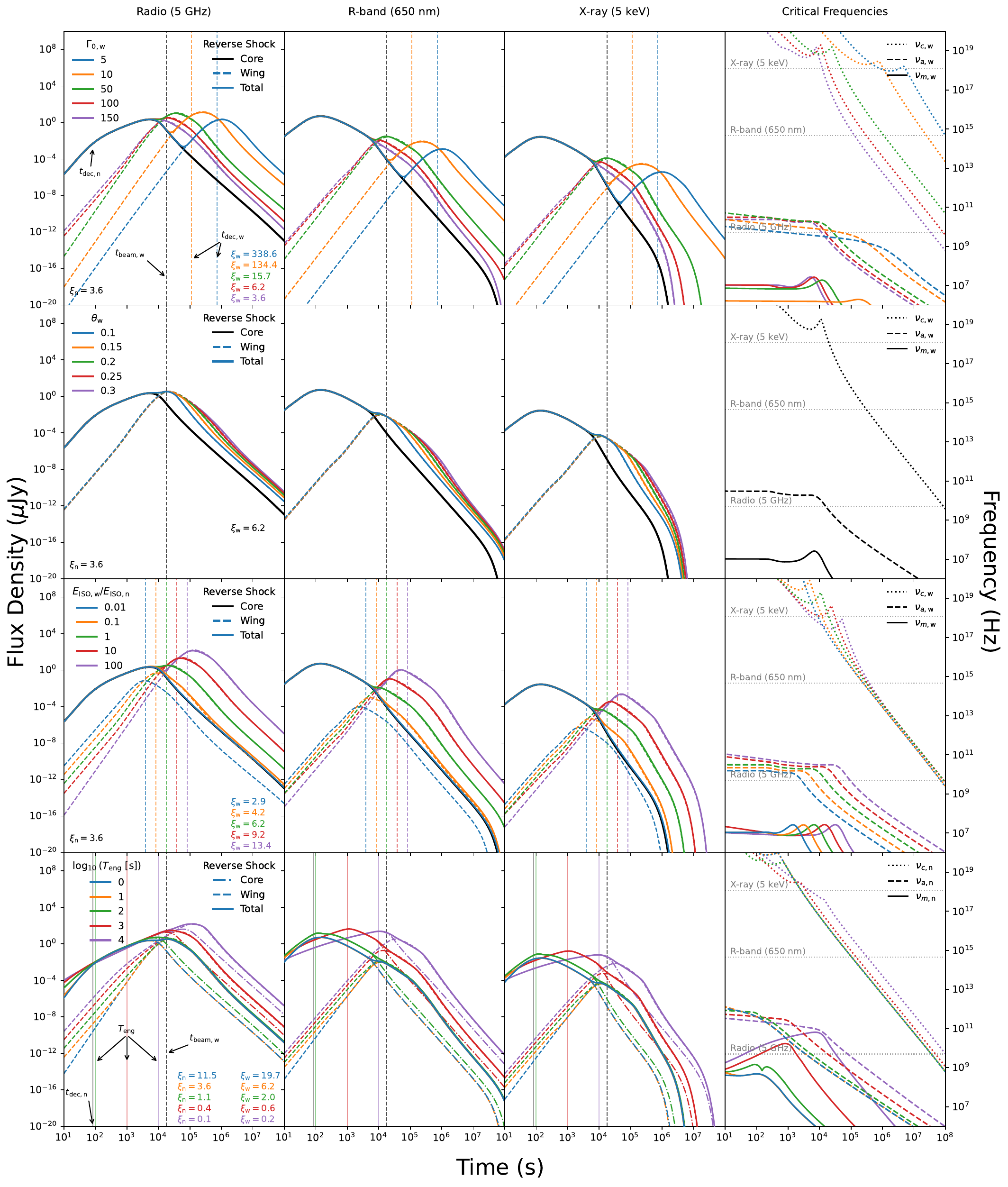}
    \caption{Multi-band light curves of reverse shock for different values of the varied parameters for the ISM. Each row corresponds to a distinct parameter, while columns correspond to different frequency bands: radio (5 GHz), R-band (650 nm), and X-ray (5 keV). The solid vertical lines in the last row depict max$(t_{\rm dec}, T_{\rm eng})$. 
    }
    \label{fig:reverse}
\end{figure*}

The reverse shock emission reaches its peak at the shock-crossing time followed by a rapid cool down, since no new electrons are accelerated \cite{Kobayashi00Light}. In the thin-shell regime, the wing-to-core ratios of the peak time and flux are given by
\begin{align}
    \mathcal{R}(t_{\rm peak}) &= \frac{1}{a(\Gamma_{0,\rm w})}\mathcal{R}(E_{\rm ISO})^{1/3} \mathcal{R}(\Gamma_0)^{-8/3}, \label{eq:RSthintime} \\
    \mathcal{R}(F_{\nu,\rm peak}) &= a(\Gamma_{0,\rm w})^\frac{p+5}{2}\mathcal{F}\mathcal{R}(E_{\rm ISO}) \mathcal{R}(\Gamma_0)^{p}. \label{eq:RSthinflux}
\end{align}
The pre-peak slopes for the core and the wing are the same, because their Lorentz factors stay constant during the shock-crossing phase. The debeaming contribution is absent. This slope is expressed as
\begin{equation}
    \frac{d\log F_\nu}{d\log t} \Bigg|_{t < t_{\rm peak}} = \frac{3(2p-1)}{2}, \label{eq:RSslope}
\end{equation}
which gives $5.4$ for $p=2.3$.

Figure~\ref{fig:reverse} shows the reverse shock light curves as a function of jet parameters.
After the core's emission reaches its peak, which happens at the time of shock crossing, it cools down quickly. This allows the rising flux from the wing to overtake. The wing is especially visible in the X-ray band, since higher frequencies reach the cutoff frequency $\nu_\mathrm{c}$ sooner \citep{Gao13complete}. Therefore, rebrightenings are more common in RS light curves compared to FS ones, provided that RS can rise above the FS emission. However, the maximum rebrightening slopes produced by the light curves in the Figure reach $t^{4.5}$. This is below the value of $t^\frac{3(2p-1)}{2}$ expected from analytical formulae, which reduces to $t^{5.4}$ for $p=2.3$ (see Eq.~\ref{eq:RSslope}). The numerical slopes are more modest because the Lorentz factor does not follow a sharp transition at $t_{\rm dec}$. Instead, $\Gamma$ begins to decrease gradually before the approximate deceleration time predicted analytically, leading to a smoother evolution and a more moderate pre-peak slope \citep{Kobayashi99Hydrodynamics}.

The first row of Fig.~\ref{fig:reverse} shows RS light curves for different initial Lorentz factors of the wing. We see that decreasing $\Gamma_{0,\rm w}$ shifts the RS peak to later times, when the core emission has already faded, making the wing contribution more pronounced. This occurs because a lower Lorentz factor corresponds to a longer shock-crossing time \citep{Sari97Hydrodynamics}, producing a more prominent RS peak compared to the rapidly cooling core emission. Thus, the RS emission from slower wings could be more easily observed, if it can rise above the FS emission.

Importantly, in the same row of Fig.~\ref{fig:reverse}, we see that the debeaming effect does not hide the dynamical peak of the RS flux, unlike in the FS case. Even when $\Gamma_{0,\rm w} > 1/\theta_{\rm n}$, the peaks occur at the shock-crossing time rather than the debeaming time. The RS flux declines steeply after its peak due to rapid cooling of the shocked ejecta. As a result, the geometric brightening associated with debeaming cannot overcome this strong decay and therefore cannot produce a separate geometric peak. However, while these dynamical peaks remain visible, their amplitudes are still suppressed by the debeaming effect. Consequently, cases with $\Gamma_{0,\rm w}>1/\theta_{\rm n}$ exhibit a reversed trend where higher Lorentz factors correspond to weaker peak fluxes. In contrast, for $\Gamma_{0,\rm w}<1/\theta_{\rm n}$, the positive correlation persists where higher $\Gamma$ results in stronger flux. Nevertheless, in all these cases the observed peak of the RS light curves corresponds to the dynamical peak. This may allow probing of $\Gamma_{0,\rm w}$ from the deceleration time $t_{\rm dec,w}$ (see Eq.~\ref{eq:tdec} in the Appendix~\ref{sec:formulas}), similar to how it was done for tophat jets \cite{Sari97Hydrodynamics, Sari99Predictions, Molinari07}. However, since the inner part of the ring is located at an angle to the observer, its arrival time will be delayed compared to the on-axis tophat case. This geometric correction should be taken into account (see Eqs.~\ref{eq:D} and~\ref{eq:Off} in the Appendix~\ref{sec:formulas}).

The second row of Fig.~\ref{fig:reverse} shows RS light curves for different values of the wing opening angle $\theta_{\rm w}$. As expected, models with smaller $\theta_{\rm w}$ exhibit earlier jet breaks, corresponding to the time when the outer edge of the wing becomes visible to the observer. Aside from this effect, we do not see a significant impact of $\theta_{\rm w}$ on the light curve peak. This is because most of the RS emission observed on-axis comes from the inner regions of the wing, so the timing of the peak is largely insensitive to the width of the outer layers. These trends mirror those seen in the FS light curves (cf. Section~\ref{sec:forward}).

In the third row of Fig.~\ref{fig:reverse}, we show the effect of different $E_{\rm ISO, w}$ on the RS light curves. Unlike FS case, we find that wings with smaller isotropic-equivalent energies can still contribute noticeably at higher frequencies, such as R and X-ray bands, because the core emission in these bands declines rapidly after its peak. Nevertheless, pronounced rebrightenings are most prevalent when $E_{\rm ISO, w} \gtrsim E_{\rm ISO, n}$, as the rising wing emission can overtake the fading core flux.

Finally, the bottom row of Fig.~\ref{fig:reverse} shows RS light curves for different central engine durations, ranging from 1 to $10^4$~s. Similarly to the forward shock, in the thin-shell regime (corresponding here to $T_{\rm eng}=1,~10,~100$~s), the shock-crossing time is primarily set by $t_{\rm dec}$, so variations in $T_{\rm eng}$ produce nearly identical peaks across all bands. The three thin-shell cases yield similar light curves, with only a minor difference for $T_{\rm eng} = 100$~s, where the shell thickness is slightly above unity ($\xi \gtrsim 1$) \citep{Wang26VegasAfterglow}. Note that \texttt{VegasAfterglow} uses "temperature"-dependent parameter $g$ that describes reverse shock dynamics. This results in lower reverse shock emission compared to analytical estimates for thin-shell case \citep{Wang26VegasAfterglow}.

In the thick-shell regime ($T_{\rm eng}=10^3,~10^4~\mathrm{s}$), the core emission reaches its peak at $T_{\rm eng}$,
\begin{equation}
    F^{\rm n}_{\nu,\rm peak} \propto E_{\rm ISO, n}^{5/4} \Gamma_{0,\rm n}^{p-2} T_{\rm eng}^{-3/4}. \label{eq:RSPeakFlux}
\end{equation}
In agreement with the equation above, we can observe the thick-shell core's R and X-ray peaks getting weaker for longer $T_{\rm eng}$. This occurs because the internal energy density of a thick shell dilutes as the crossing time is pushed to larger radii. Conversely, the radio peak grows stronger because it is dictated by the spectral transition as the characteristic absorption frequency $\nu_\mathrm{a}$ crosses the band. The delayed transition in thicker shells allows the radio flux to reach a higher level before the material becomes transparent at that frequency.

The wing's peak in the thick-shell regime is also set by $T_{\rm eng}$. However, the wing's inner edge is at an angle to the observer and its arrives later compared to the core emission. The wing's shock-crossing time and flux in the observer's frame are given by
\begin{align}
    t_{\times,\rm w} &= T_{\rm eng} \times a(\Gamma_{\rm w}(t_{\times,\rm w}))^{-1}, \\
    F_{\nu, \rm peak} &\propto a(\Gamma_{\rm w}(t_{\times,\rm w}))^{\frac{p+5}{2}} \mathcal{F} E_{\rm ISO, w}^{5/4} \Gamma_{0,\rm w}^{p-2} T_{\rm eng}^{-3/4}.
\end{align}
Unlike core, the thick-shell wing's R and X-ray peaks become stronger for longer engine durations despite the $\propto T_{\rm eng}^{-3/4}$ proportionality in the equation above. This trend is caused by the factor $a(\Gamma_{\rm w}(t_{\times,\rm w}))^{\frac{p+5}{2}}$ which encompasses the flux suppression due to it being beamed away. Because a larger $T_{\rm eng}$ pushes the shock-crossing moment to later times when the wing has decelerated further, the resulting lower $\Gamma_{\rm w}(t_{\times,\rm w})$ leads to a less suppressed--and thus stronger--observed flux.

Additionally, we observe that the wing flux preserves a steep pre-peak slope even in the thick-shell regime, in contrast to the core emission. As in the FS case, the debeaming effect introduces an additional geometric contribution to the temporal slope, such that
\begin{equation}
    \frac{d\log F_\nu^{\rm w}}{d\log t} \Bigg|_{t < t_{\rm peak}} = \frac{1}{2} + \frac{\Gamma_{\rm w}^2\theta_{\rm n}^2}{2 + \Gamma_{\rm w}^2\theta_{\rm n}^2} \left(\frac{p}{2} + 3\right), \label{eq:RSslopethick}
\end{equation}
which approaches a maximum value of $\simeq4.6$ for $p=2.3$ in the limit $\Gamma_{\rm w}\theta_{\rm n} \gg 1$. The first term reflects the intrinsic dynamics in the thick-shell regime, while the second term arises from the time dependence of the Lorentz factor and dominates when the wing is strongly beamed away from the observer.

\subsection{Wind Medium}
\label{sec:wind}

\begin{figure*}[p]
    \centering
    \includegraphics[width=\textwidth]{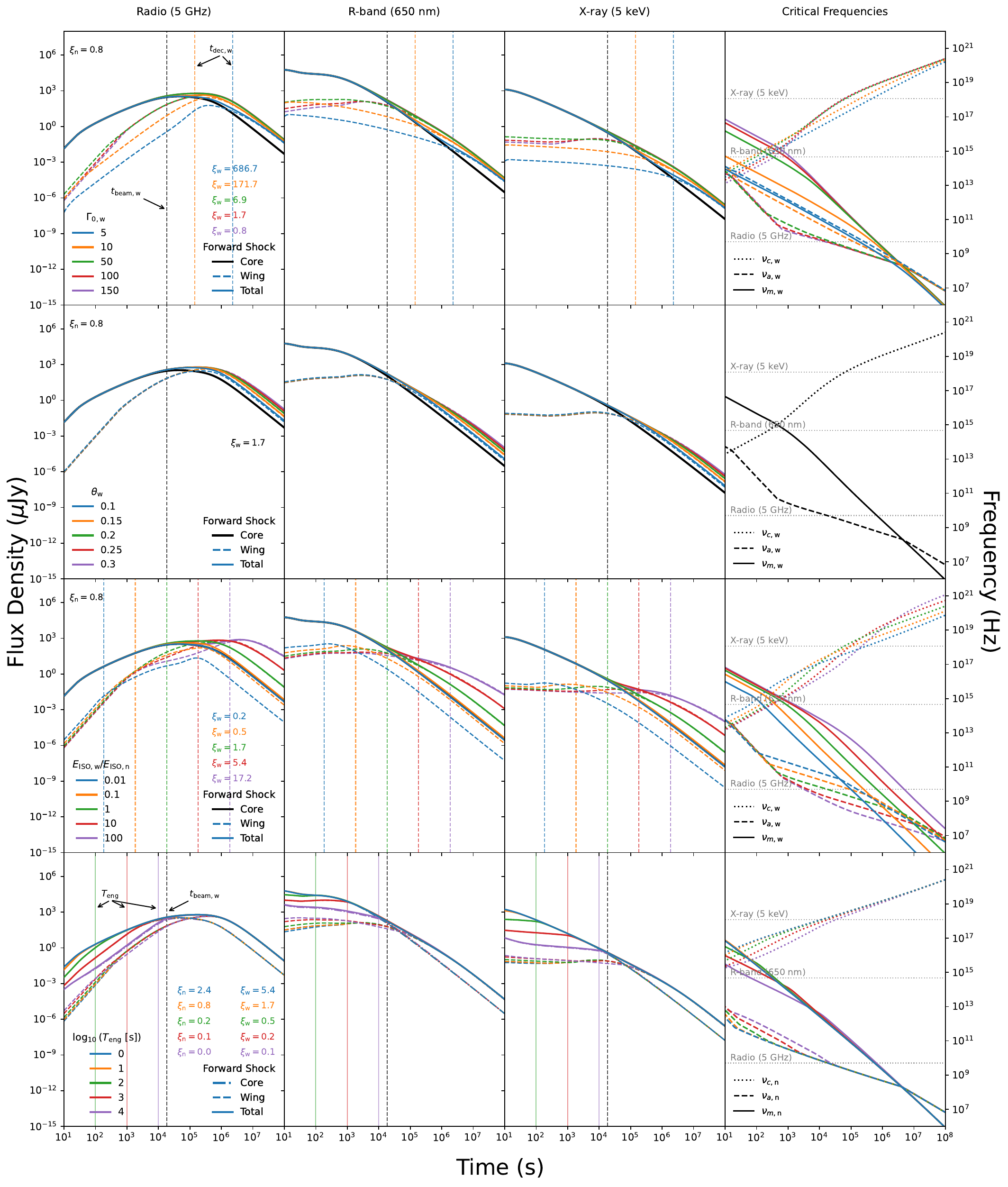}
    \caption{Multi-band light curves of forward shock for different values of the varied parameters in the wind medium. Each row corresponds to a distinct parameter, while columns correspond to different frequency bands: radio (5 GHz), R-band (650 nm), and X-ray (5 keV). The solid vertical lines in the last row depict max$(t_{\rm dec}, T_{\rm eng})$. The core deceleration time is outside the range, $t_{\rm dec, n}\approx6~\rm{s}$.}
    \label{fig:wind_forward}
\end{figure*}

\begin{figure*}[p]
    \centering
    \includegraphics[width=\textwidth]{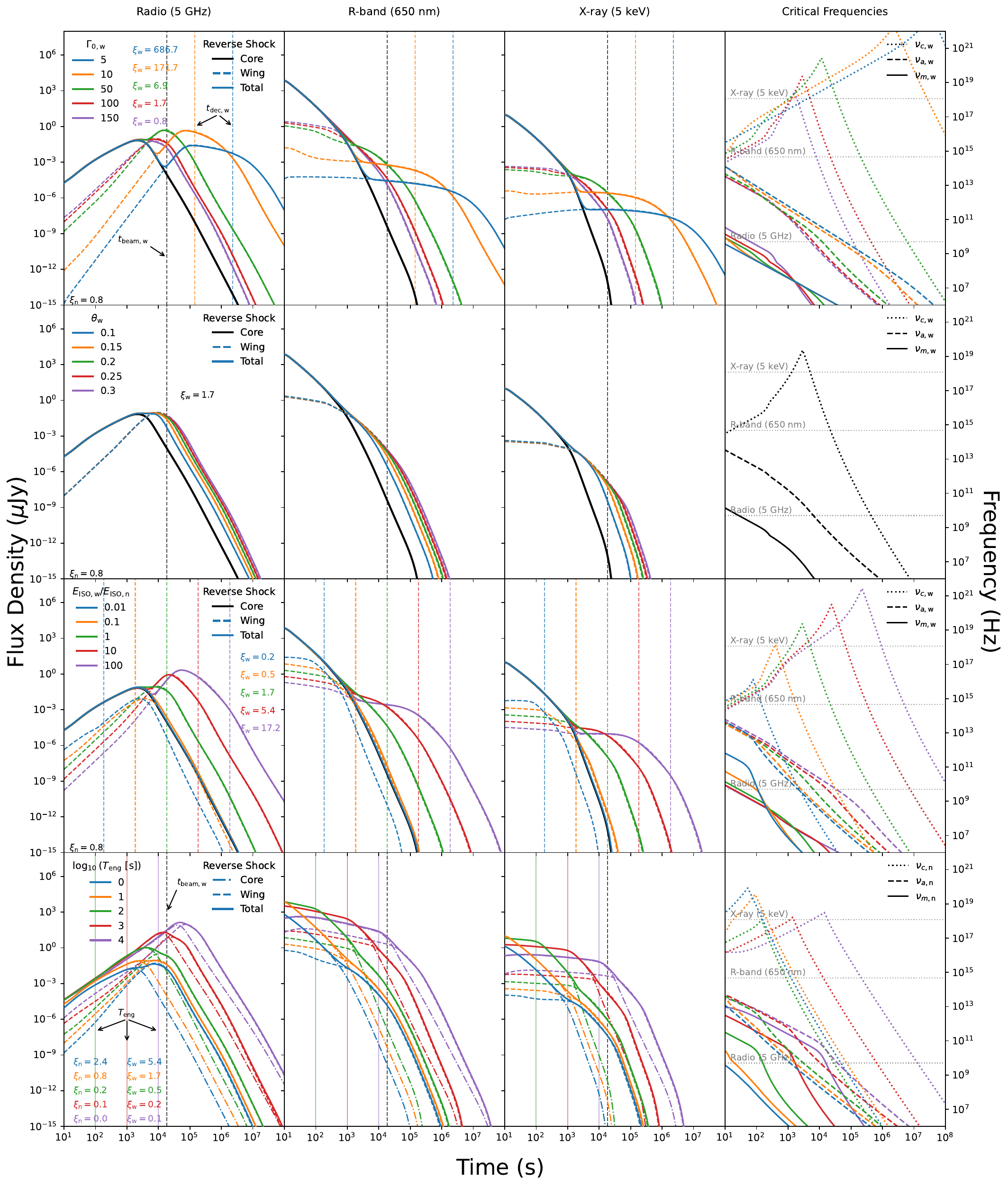}
    \caption{Multi-band light curves of reverse shock for different values of the varied parameters in the wind medium. Each row corresponds to a distinct parameter, while columns correspond to different frequency bands: radio (5 GHz), R-band (650 nm), and X-ray (5 keV). The solid vertical lines in the last row depict max$(t_{\rm dec}, T_{\rm eng})$. The core deceleration time is outside the range, $t_{\rm dec, n}\approx6~\rm{s}$. }
    \label{fig:wind_reverse}
\end{figure*}

In this section, we discuss the wind medium case. Our discussion follows the pattern similar to the ISM case above. Fig.~\ref{fig:wind_forward} and \ref{fig:wind_reverse} show FS and RS light curves, respectively, in a wind-like medium. Each figure presents variations of the same key parameters: the initial wing Lorentz factor $\Gamma_{0,\rm w}$, the wing opening angle $\theta_{\rm w}$, the isotropic-equivalent energy $E_{\rm ISO, w}$, and the central engine duration $T_{\rm eng}$. Similarly, each parameter was changed separately while others were fixed at their base values on Table~\ref{tab:const_params}.

The first rows of Fig.~\ref{fig:wind_forward} and ~\ref{fig:wind_reverse} illustrate the effect of varying the initial wing Lorentz factor. For the forward shock, the R and X-ray light curves decline monotonically \citep{Panaitescu00Analytic}. Highly relativistic wings ($\Gamma_{0,\rm w} > 20$) exhibit a mild steepening of the flux slopes during the shock-crossing phase. This is most clearly observed in the radio and X-ray bands. The debeaming effect enhances the slopes by a smaller extent compared to the ISM case. This is associated with the slower post-crossing deceleration in a wind profile ($\Gamma \propto t^{-1/4}$ versus $t^{-3/8}$ in ISM), which weakens the debeaming contribution to slopes. For slower wings ($\Gamma_{0,\rm w} < 20$), which are within the observer's field of view from the start, the behavior resembles the ISM case in that lower Lorentz factors correspond to weaker fluxes. Overall, the dependence on $\Gamma_{0,\rm w}$ follows the qualitative trends found in the ISM case.

The RS emission in the wind medium also follows a similar pattern as in the ISM. Slower wings have longer deceleration times and therefore produce RS emission at later epochs \citep{Kobayashi03Early, Yi13Early}, when the core component has already cooled significantly. However, a difference from the ISM case can be observed in the highly relativistic cases ($\Gamma_{0,\rm w}>20$), where R and X-ray flux strengths maintain a positive correlation with $\Gamma_{0,\rm w}$. This reflects a stronger dependence of flux on the Lorentz factor compared to the ISM case. Higher values of $\Gamma_{0,\rm w}$ produce stronger emissions that compensate for the beaming suppression.

In the second rows of Figs.~\ref{fig:wind_forward}-\ref{fig:wind_reverse}, we demonstrate the impact of varying $\theta_{\rm w}$. For both the forward and reverse shocks, narrower wings produce earlier jet breaks, corresponding to the time when the outer edge of the ring becomes visible to the observer. Aside from this geometric effect, the light curves remain insensitive to $\theta_{\rm w}$. This is because the dominant on-axis emission originates from the inner regions of the wing. This behavior mirrors the result found in the ISM case above.

The third rows of Figs.~\ref{fig:wind_forward}-\ref{fig:wind_reverse} correspond to the variation of the wing isotropic-equivalent energy. Similarly to the ISM case, low-energy wings remain subdominant and hidden by the core emission, while sufficiently energetic wings can dominate the total flux at late times. In the R and X-ray bands, however, this dominance does not produce rebrightenings in the wind-like medium. Instead, the light curves remain monotonically decaying. The observable jet break corresponds to the wing component, while the jet break of the core becomes concealed. For the reverse shock, the rapid decline of the core emission at high frequencies makes the wing contribution more visible, but again without pronounced rebrightening outside the radio band. In contrast, the radio band reveals the wing component in both the FS and RS, as the spectral peak--occurring when $\nu_\mathrm{a}$ crosses the observed frequency--appears later than that of the core \citep{Zou05Early}. In the RS case, the rising radio flux can produce pre-peak slopes as steep as $t^3$, as inferred from our numerical light curves, when the self-absorption frequency $\nu_\mathrm{a}$ approaches the observed band. This slope is steeper than the analytically estimated $t^{2.5}$ in the thin-shell case (see Fig.~\ref{fig:Wind_RS_thin_tikz}). A primary cause of this discrepancy lies in the microphysical treatment of the injected electrons. Standard analytical solutions for a thin-shell reverse shock assume the accelerated electrons are at least mildly relativistic ($\gamma_m \gg 1$), while the numerical framework properly accounts for the deep-Newtonian regime ($\gamma_m\approx1$) throughout the whole evolution. Consequently, the numerical models apply a suppression factor, $f_{\rm syn}\propto (\gamma_m-1)$, to the emission because the synchrotron radiation is less efficient. This correction factor boosts the pre-peak slope by $(\gamma_m-1)\propto t$. This applies only to the thin-shell RS emission, because in thick shells the shock becomes relativistic early.

The bottom rows of Fig.~\ref{fig:wind_forward} and \ref{fig:wind_reverse} show the FS and RS light curves, respectively, for different values of the engine duration. Within our chosen parameter space, the wind-like medium shows a higher prevalence of thick shells compared to the ISM case. Consequently, most of the models present here fall into thick or moderately thin regimes ($\xi \gtrsim 1$). The only clearly thin-shell case corresponds to $T_{\rm eng}=1$~s. Unlike the ISM case, the Lorentz factor evolves similarly in thin and thick shells in the wind medium (cf. Fig.~\ref{fig:Gamma_dur}). In the thick-shell regime, $\Gamma$ rapidly decreases from $\Gamma_0$ at very early times to a constant value near the critical Lorentz factor
\begin{equation}
    \Gamma_{\rm c} \propto \left(\frac{E_{\rm ISO}}{A_* T_{\rm eng}}\right)^{1/4}
\end{equation}
well before the shock-crossing time. It then remains constant throughout the crossing phase and begins to decelerate only afterward in the self-similar regime. As a result, thick shells evolve similarly to the thin shells, and there is no geometric slope enhancement associated with debeaming in either thin- or thick-shell regime. This contrasts with the ISM case, where in the thick-shell regime $\Gamma$ is both closer to $\Gamma_0$ and continuously decreasing during the crossing, producing stronger Doppler suppression and an additional slope enhancement from active deceleration.

For the forward shock, the core emission during shock crossing becomes progressively weaker as $T_{\rm eng}$ increases, since the injected energy is spread over a longer duration. The wing, however, shows the opposite trend. As discussed above, in the wind case the thick-shell Lorentz factor settles to a lower value prior to shock crossing. For longer $T_{\rm eng}$, this value is smaller, which reduces the Doppler suppression of the wing emission. Consequently, the wing's early flux increases with $T_{\rm eng}$, while the core emission weakens.

The reverse shock shows trends broadly similar to the ISM case. In the thick-shell regime, the core R and X-ray peaks decrease with increasing $T_{\rm eng}$ due to energy dilution, while the radio peak becomes stronger and shifts to later times because the self-absorption frequency $\nu_\mathrm{a}$ crosses the band at a later epoch. The wing behaves analogously to the ISM case: in the R and X-ray bands, longer $T_{\rm eng}$ leads to stronger observed flux, since the wing Lorentz factor is smaller and the geometric suppression is reduced (see Fig.~\ref{fig:Gamma_dur}).

\subsection{Wind-to-ISM transition}
\label{sec:wind2ism}

\begin{figure*}[t]
    \centering
    \begin{subfigure}[t]{0.45\textwidth}
        \centering
        \includegraphics[width=\linewidth]{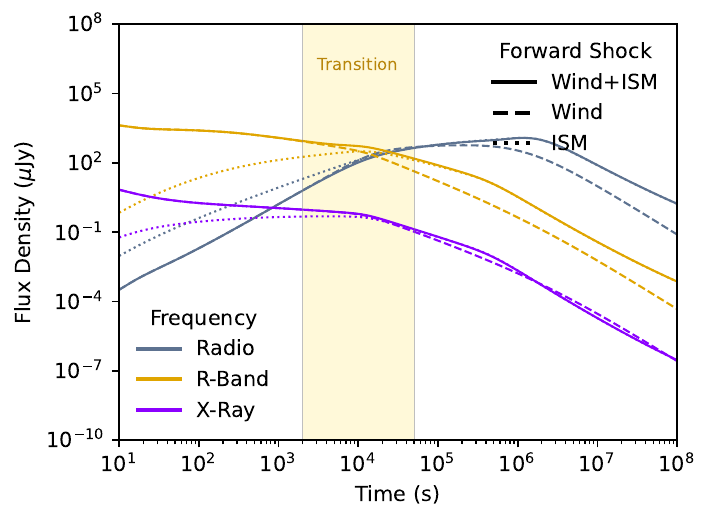}
    \end{subfigure}
    \begin{subfigure}[t]{0.45\textwidth}
        \centering
        \includegraphics[width=\linewidth]{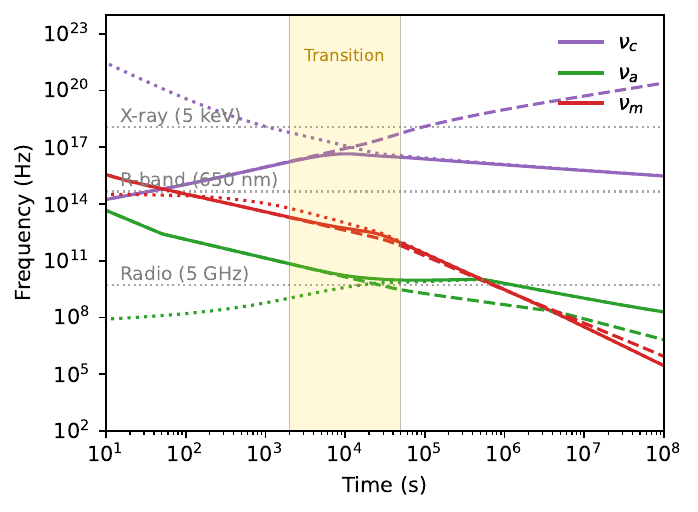}
    \end{subfigure}
    \begin{subfigure}[t]{0.45\textwidth}
        \centering
        \includegraphics[width=\linewidth]{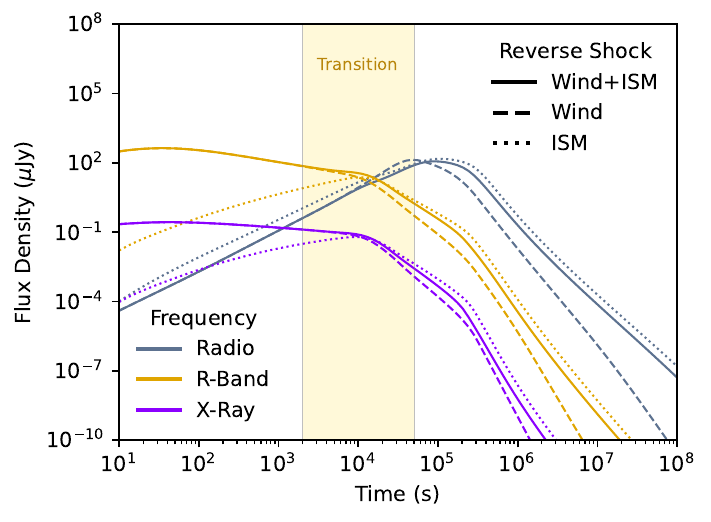}
    \end{subfigure}
    \begin{subfigure}[t]{0.45\textwidth}
        \centering
        \includegraphics[width=\linewidth]{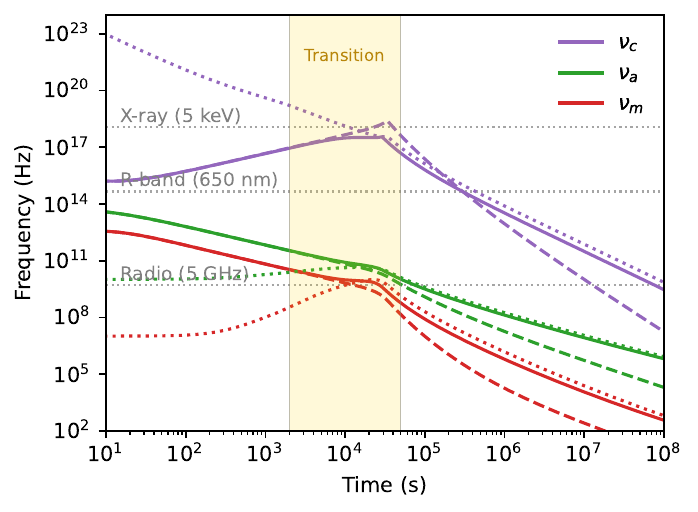}
    \end{subfigure}
    \caption{Multi-band light curves and critical frequencies for forward and reverse shocks in the stratified medium with wind-to-ISM transition for $A_{\rm star}=0.1$, $n_{\rm ISM}=1$~cm$^{-3}$, and $T_{\rm eng}=10^4$~s.}
    \label{fig:stratified}
\end{figure*}

GRB jets that initially propagate in a stellar-wind environment may eventually encounter the surrounding ISM. Observational evidence for such transitions has already been reported \citep[e.g.,][]{Li20GRB140423A, Fraija17Theoretical}. We therefore consider a stratified external medium that initially follows a wind density profile and gradually transitions to a homogeneous ISM.

In \texttt{VegasAfterglow}, this transition is modeled in a simplified manner: once the wind density declines to the ambient ISM density, the profile smoothly approaches a constant-density medium. Physically, however, the wind–ISM interface is expected to produce a density discontinuity, along with forward and reverse shocks and the corresponding density jumps \citep{Dai03GRB030226, Peer06Signature, vanEerten09No, Peer24Gamma}. Despite this simplification, we use this method to highlight the key aspects of the signatures of the wind-to-ISM transition for two-component jets. 

Figure~\ref{fig:stratified} shows the light curves for such a medium. In this example, we assume $n_\mathrm{ISM}=1 \, \mathrm{cm}^{-3}$, $A_*=0.1$, and $T_\mathrm{eng}=10^4 \, \mathrm{s}$. The choice of a longer engine duration differs from the default value adopted in this paper ($T_{\rm eng}=10~\mathrm{s}$). This is done to ensure that the reverse-shock emission remains significant at the time when the density profile transitions from wind to ISM. For shorter engine durations, the reverse shock fades before the transition occurs, making the effect of the stratified medium on the light curves medium difficult to identify.

The figure is arranged in two rows and two columns. The top row shows the forward shock: light curves are shown on the left panel and the corresponding critical frequencies are shown on the right panel. The bottom row presents the same quantities for the reverse shock. In each panel, we compare three cases in the radio, R, and X-ray bands: pure wind (dashed), pure ISM (dotted), and the stratified wind+ISM profile (solid). The shaded region marks the transition interval, defined as the time range over which the density deviates by more than 1\% from either the pure wind or pure ISM solution.

At early times, before the transition region, the stratified light curves coincide exactly with the pure wind solution for both FS and RS. In this phase, the R and X-ray fluxes are stronger than in the ISM case, while the radio flux is weaker, fully consistent with the wind-like density profile (cf. Section~\ref{sec:wind}). Moreover, throughout most of the transition interval, the light curves and spectral evolution continue to closely follow the wind behavior. Noticeable deviations appear only toward the end of the transition phase.

Near the end of the transition phase, the pure wind and pure ISM solutions intersect, with both the light curves and their corresponding critical frequencies reaching comparable values. Although the exact intersection shows minor band-to-band variations, the crossings occur within a narrow temporal window. Beyond this point, the stratified solution converges to the ISM behavior, fully reproducing the ISM-like evolution for both FS and RS emission. The spectral transitions now follow the ISM ordering, leading to ISM-type peaks in the radio and R-band and cooling behavior in the reverse shock. Overall, the stratified density profile produces an interpolation between wind and ISM regimes, with the emission closely tracking whichever density profile dominates at a given time.

\subsection{Circumburst medium density}
\label{sec:density}

\begin{figure*}[p]
    \centering
    \includegraphics[width=\textwidth]{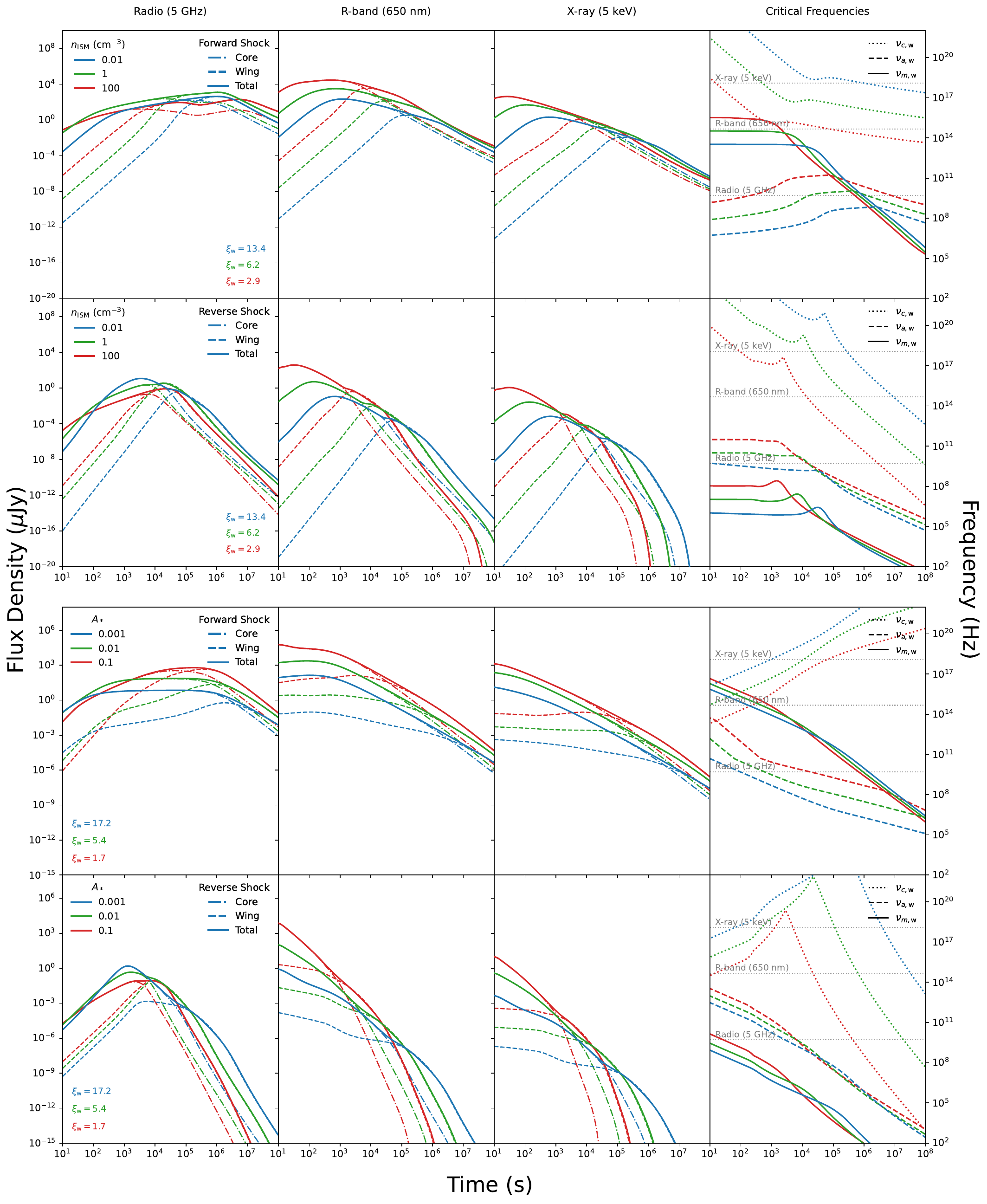}
    \caption{Multi-band light curves for forward and reverse shocks in the ISM and wind media for different values of $n_{\rm ISM}$ and $A_{\rm star}$.}
    \label{fig:densities}
\end{figure*}

The dependence of afterglow light curves on the circumburst medium density has been extensively studied in the literature \citep[e.g.,][]{Sari98Spectra, Chevalier99Gamma, Chevalier00Wind, Granot02Shape, Tian22Constraining}. For completeness, here we briefly summarize the expected systematics in the case of two-component jets.

Figure~\ref{fig:densities} illustrates how the light curves depend on the external density in the ISM and the wind-like environments. The first and second rows correspond to the forward and reverse shock in the ISM, respectively. Similarly, the third and fourth rows correspond FS and RS for the wind-like medium. In each row, we show radio, R-band, and X-ray light curves, as well the wing critical frequencies while varying the density parameter: $n_{\rm ISM} = 0.01, 1, 100~\mathrm{cm^{-3}}$ for the ISM case, and $A_*=0.001, 0.01, 0.1$ for the wind case.

The first row presents the FS emission in the ISM. As expected, increasing $n_{\rm ISM}$ raises the self-absorption frequency $\nu_\mathrm{a}$ and lowers the cooling frequency $\nu_\mathrm{c}$, while the injection frequency $\nu_\mathrm{m}$ attains higher values before the deceleration time but subsequently begins its decline earlier due to the shorter $t_{\rm dec}$ \citep[][]{Gao13complete}. The overall flux normalization increases and the deceleration time $t_{\rm dec}$ becomes shorter, shifting the dynamical peaks to earlier times and higher fluxes \cite{Sari97Hydrodynamics}. This behavior is evident in the wing's R and X-ray peaks. In contrast, the core's R-band produces a spectral peak when the observing frequency crosses $\nu_\mathrm{m}$. The timing of this crossing is largely insensitive to the external density, so the core R-band peaks occur at approximately the same time for different $n_{\rm ISM}$.

Moreover, a given observing band may occupy different spectral regimes depending on the density. For example, the wing's radio band remains above $\nu_\mathrm{a}$ for $n_{\rm ISM}=0.01~\mathrm{cm^{-3}}$, whereas for higher densities the self-absorption frequency can exceed the radio band, producing a distinctive peak later when $\nu_\mathrm{a}$ drops and crosses the band again.

The same qualitative behavior is seen for the reverse shock in the ISM, as shown in the second row of Fig.~\ref{fig:densities}. Increasing $n_{\rm ISM}$ enhances the emission strength and shifts the dynamical evolution to earlier times. The characteristic frequencies $\nu_\mathrm{a}$ and $\nu_\mathrm{m}$ reach higher values during the shock-crossing phase, but begin their decline earlier due to the shorter deceleration time, while $\nu_\mathrm{c}$ decreases overall. As a result, the dynamical peaks move to earlier times and higher flux levels, consistent with the forward-shock behavior \citep{Kobayashi03Early}.

In the R and X-ray bands--both remaining above $\nu_\mathrm{m}$ and $\nu_\mathrm{a}$ throughout the evolution--the flux increases systematically with density. In contrast to the forward-shock case, the core's R-band peak is dynamical rather than spectral and therefore follows the same density dependence as the wing's peak. The radio emission, however, does not exhibit a simple monotonic trend, since spectral transitions modify the light curves and partially obscure the underlying density scaling.

Similarly, in the third and fourth rows we observe the effect of the wind density parameter $A_*$ on the light curves and critical frequencies. As in the ISM case, increasing $A_*$ enhances the overall flux normalization, raises $\nu_\mathrm{a}$ and $\nu_\mathrm{m}$, and lowers $\nu_\mathrm{c}$ with $t_{\rm dec}$ \citep[][]{Gao13complete}. For the forward shock, the wind medium does not produce dynamical peaks. Instead, the core R-band exhibits a spectral peak associated with the crossing of the injection frequency $\nu_\mathrm{m}$ \citep{Zou05Early}, whose timing remains approximately insensitive to $A_*$. The X-ray band, lying above $\nu_\mathrm{m}$ throughout the evolution, shows no distinct peak for either component and simply increases in normalization with density. The radio band clearly reflects the spectral evolution: the frequency crosses $\nu_\mathrm{a}$ and reaches a stronger flux for higher $A_*$ until it crosses $\nu_\mathrm{m}$.

For the reverse shock, both the R and X-ray bands of the core and wing remain above $\nu_\mathrm{a}$ and $\nu_\mathrm{m}$, and therefore decay monotonically while increasing in normalization with higher $A_*$. The RS radio emission, however, does not display a simple monotonic trend, as different spectral transitions take place for different densities and obscure a clear dependence on $A_*$, similar to the ISM case.

\subsection{Total Flux}
\label{sec:total}

\begin{figure*}[t]
    \centering
    \includegraphics[width=\textwidth]{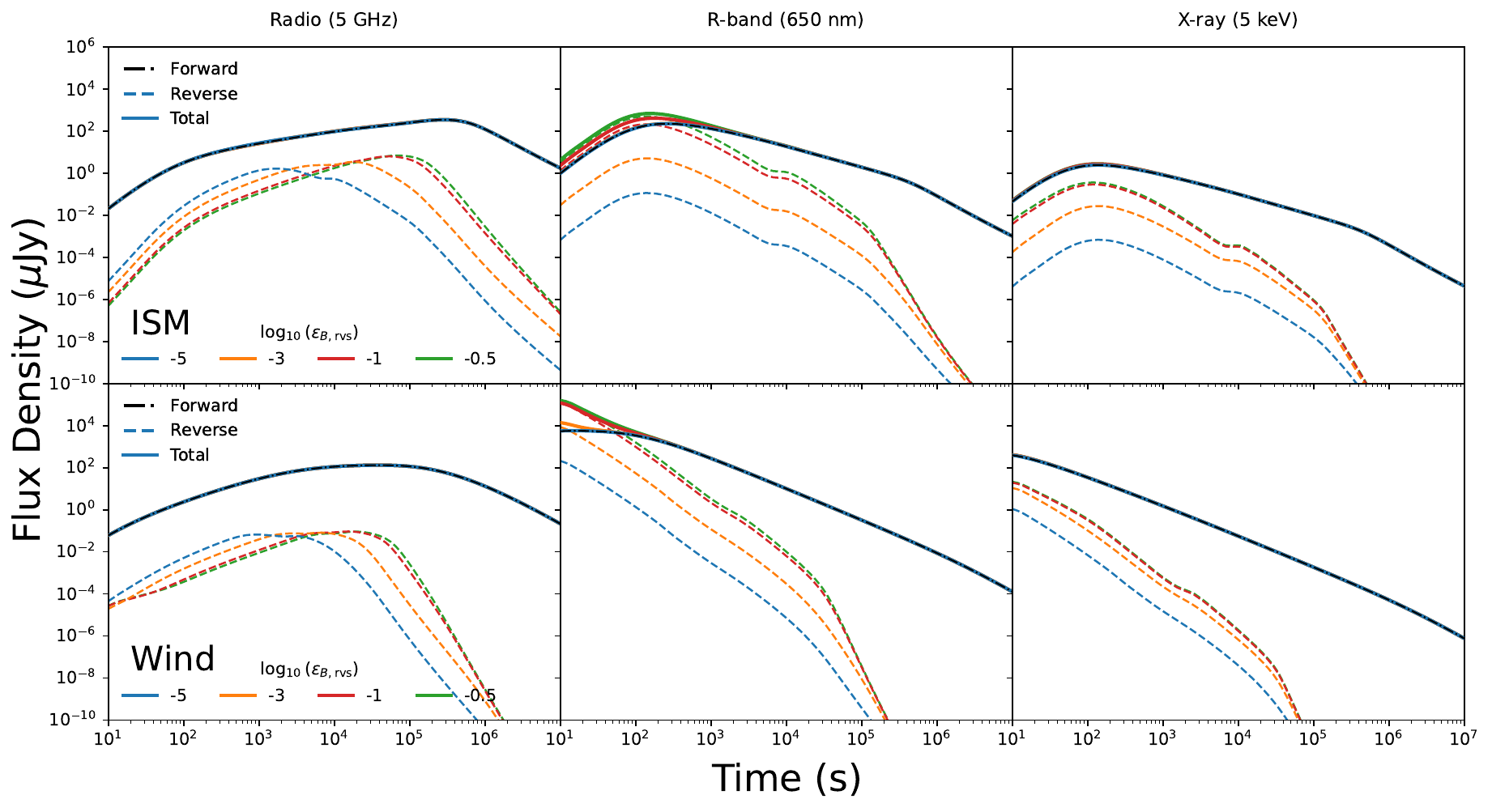}
    \caption{Multi-band light curves of forward and reverse shocks combined for different values of $\varepsilon_{B,\rm rvs}$ in the ISM and wind media. Each column corresponds to different frequency bands: radio (5 GHz), R-band (650 nm), and X-ray (5 keV). The default parameters are the same as in Table~\ref{tab:const_params} with the exception of $\varepsilon_{B,\rm fwd}=10^{-5}$.}
    \label{fig:total}
\end{figure*}

\begin{figure*}[t]
    \centering
    \includegraphics[width=\textwidth]{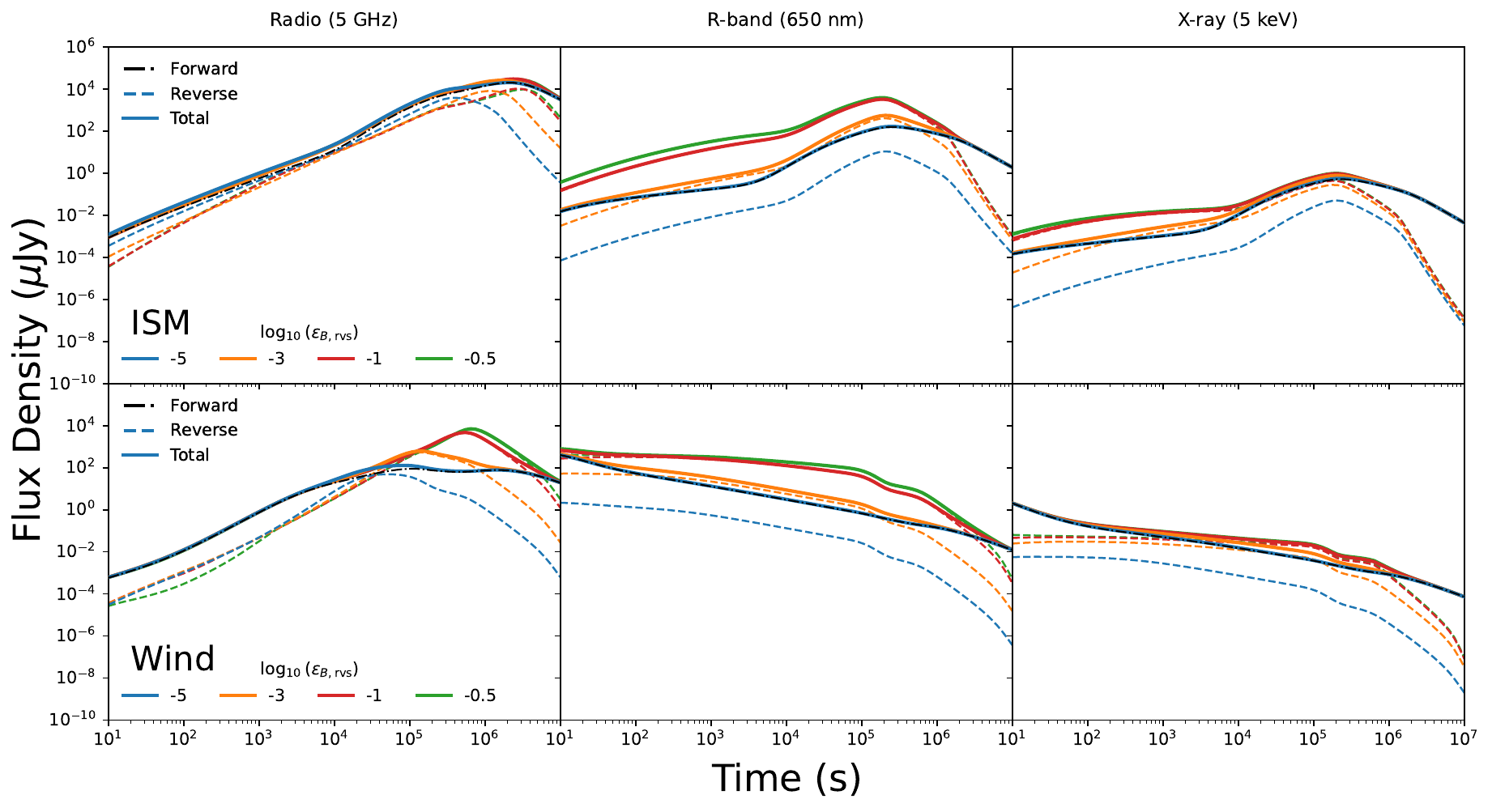}
    \caption{Multi-band light curves of forward and reverse shocks combined for different values of $\varepsilon_{B,\rm rvs}$ in the ISM and wind media. Each column corresponds to different frequency bands: radio (5 GHz), R-band (650 nm), and X-ray (5 keV). The default parameters are the same as in Table~\ref{tab:const_params} with the exception of $\varepsilon_{B,\rm fwd}=10^{-5}$, $E_{\rm ISO,w}/E_{\rm ISO,n}=100$, and $T_{\rm eng} = 10^5$~s.}
    \label{fig:total_2}
\end{figure*}

The total flux is a sum of emission not only from core and wing of the jet, but also from forward and reverse shocks. While the relative contribution of RS emission has been studied extensively in the literature \citep[e.g.,][]{Sari99Predictions, Zhang05Gamma, Japelj14RS, Gao15Reverse, Yi20Bright}, here we briefly highlight aspects that are relevant for two-component jets. 

To assess the role of the reverse shock in the observable signal, we study the total emission obtained by summing the FS and RS contributions, while varying only the reverse-shock magnetization parameter $\varepsilon_{B,\rm rvs}$. We fix $\varepsilon_{e,\rm fwd} = \varepsilon_{e,\rm rvs} = 10^{-1}$ \citep{Nava2014LAT,Beniamini17epsilone} and adopt a reduced forward-shock magnetization, $\varepsilon_{B,\rm fwd} = 10^{-5}$ (lower than the fiducial $10^{-3}$ although consistent with afterglow studies \citep{Santana2014epsilonB,Beniamini2015energies}), in order to enhance the relative visibility of the reverse-shock component.

We consider two setups. In the first, we adopt the default parameters listed in Table~\ref{tab:const_params}, including a short engine duration $T_{\rm eng}=10~\mathrm{s}$. In the second, we construct an extreme configuration designed to maximize the reverse-shock contribution, with $T_{\rm eng}=10^5~\mathrm{s}$ and an enhanced wing energy contrast $E_{\rm ISO, w}/E_{\rm ISO,n}=100$. All other parameters are kept unchanged.

Figures~\ref{fig:total} and~\ref{fig:total_2} show the combined light curves for $\varepsilon_{B,\rm rvs} = 10^{-5},~10^{-3},~10^{-1},~10^{-0.5}$. The first figure corresponds to the first setup, while the second illustrates the extreme case. On each figure, the first row corresponds to the ISM case and the second row to the wind medium, showing radio, R, and X-ray bands.

In the default configuration, the forward shock dominates the total emission during most of the period. In the radio and X-ray bands, the RS contribution remains subdominant at all times, including the extreme case $\varepsilon_{B,\rm rvs}=10^{-0.5}$, despite the large contrast with $\varepsilon_{B,\rm fwd}$. In the R-band, the core RS emission during the shock-crossing phase can produce a visible bump, particularly for $\varepsilon_{B,\rm rvs}=10^{-1},~10^{-0.5}$. At later times, although the wing overtakes the core in the RS component, the total emission is already dominated by the FS, rendering the wing's reverse-shock contribution effectively unobservable. These trends are present in both ISM and wind media, with the early R-band bump slightly more pronounced in the ISM case.

In contrast, the extreme configuration leads to a different behavior, with the reverse shock becoming comparable to or even dominating the forward shock over extended periods. The effect is most pronounced in the R-band, where the RS exceeds the FS for most of the evolution. In the ISM case particularly, the wing's RS rebrightening is clearly visible for $\varepsilon_{B,\rm rvs}=10^{-1},~10^{-0.5}$.

In the radio band, the behavior depends on the external medium. In the ISM, the RS and FS remain comparable, whereas in the wind medium the wing RS produces a prominent late-time peak, leading to a detectable rebrightening.

In X-rays, the RS remains subdominant overall, but the dominant component differs between environments. In the ISM, the core's RS emission dominates early on, eventually becoming hidden behind the wing's FS flux. In the wind medium, the trend is reversed: the core FS dominates at early times and the wing RS can briefly emerge at late times before fading again.

Overall, these results show that the FS remains dominant for standard parameter choices, and observable RS signatures are generally limited to early-time optical features associated with the core. However, in extreme scenarios involving long-lived engines and strongly energetic wings, the reverse shock---from both the core and the wing---can leave detectable imprints on the emission.

\section{Conclusion}
\label{sec:conclusion}

We have conducted a systematic exploration of broadband afterglow emission from two-component gamma-ray burst (GRB) jets. Using numerical modeling with the {\tt VegasAfterglow} code complemented by analytical scalings, we quantified the dependence of the emission on key jet parameters, including the initial wing Lorentz factor, wing opening angle, energy distribution, and central engine duration. By investigating uniform (ISM), stellar wind, and stratified environments across thin- and thick-shell regimes, we mapped the physical conditions required for the wing component to produce observable signatures for on-axis observers.

The wing emission is effectively off-axis, so its visibility depends on its initial Lorentz factor. For $\Gamma_{0,\rm w}>\theta_{\rm n}^{-1}$, the wing is initially beamed away from the observer and only becomes visible after deceleration. This produces a debeaming effect that steepens the rising light-curve slopes, reaching up to $\sim t^{4.5}$ in our simulations, while not affecting the late-time decline after the jet break. As a result, this model is unlikely to account for rapidly declining transients with decay slopes steeper than $t^{-p}$ \citep[e.g.,][]{Sfaradi25First}, which represents the steepest decline typically expected after a jet break \cite{Sari99Jets}. In the forward shock, this steepening can obscure the dynamical peak associated with the shock-crossing time and instead produce a geometric peak when $\Gamma_{\rm w}\sim\theta_{\rm n}^{-1}$, erasing the signature of the wing's initial Lorentz factor. This typically occurs near the core jet-break time, when the core's outer edge and the wing's inner edge enter the observer's line of sight. In contrast, the reverse-shock emission declines rapidly after shock crossing, preventing the formation of a geometric peak and allowing the dynamical peak to remain distinguishable even for highly relativistic wings.

Producing a distinct late-time rebrightening requires the wing to carry sufficient energy. For the forward shock, pronounced peaks arise only if the wing is substantially more energetic than the isotropic-equivalent energy of the core. In contrast, for the reverse shock, well-separated shock-crossing times allow even comparable isotropic-equivalent energies to generate clearly identifiable bumps. The rebrightening typically occurs at $10^4-10^5\,\mathrm{s}$ after the burst, shortly after the core jet-break time. In a homogeneous ISM, this feature is associated with the wing's rising flux during the shock crossing time and it appears in the optical and X-ray bands. In contrast, in a wind environment, dynamical rebrightenings are absent due to the lack of pronounced peaks in the light curves. Instead, any late-time rebrightening is of spectral origin, occurring in the radio band as the self-absorption frequency crosses the band.

When considering jets propagating through a stratified medium that transitions from a stellar wind to a homogeneous ISM, the afterglow smoothly interpolates between the pure wind and pure ISM solutions. The emission mostly tracks the locally dominant density profile at any given time.

Variations in the external density primarily rescale the afterglow evolution without introducing qualitatively new phenomenology. In both ISM and wind-like environments, increasing the density enhances the overall flux normalization and shifts the dynamical evolution to earlier times, so that peaks--when present--occur earlier and at higher flux levels. R-band and X-ray emission typically varies monotonically with density, while radio light curves can exhibit more complex evolution due to spectral transitions that partially obscure the underlying scaling with density.

Our study of the combined forward and reverse shock emission shows that the total flux is dominated by the forward shock for most of the parameter space, with the reverse shock becoming observable under specific conditions. In particular, the RS magnetization should be at least two orders of magnitude larger than the RS magnetization, $\varepsilon_{B,\rm rvs} \gtrsim 10^2\varepsilon_{B,\rm fwd}$. In this scenario, the reverse shock can dominate the early optical emission, peaking at the shock-crossing time $t_\times$ and remaining visible until $\sim 10\,t_\times$ before being overtaken by the forward shock. A similar but weaker trend is present in X-rays, where the reverse shock contribution is less prominent. In contrast, the radio RS emission is typically suppressed at early times and emerges only at late times ($t \gtrsim t_\times$), as the self-absorption frequency approaches the observing band. The wing's contribution to the RS becomes appreciable only for sufficiently large engine durations ($T_{\rm eng}\gtrsim10^4\,\mathrm{s}$), where the wing RS component has sufficient time to overtake the core RS flux before the total emission becomes dominated by the forward shock.

The results also have important implications for energy estimates. The prompt emission is dominated by the jet core, while the late-time afterglow can be dominated by the wing. Consequently, the isotropic-equivalent energy inferred from the prompt gamma-ray energy, $E_{\rm ISO,\gamma}$, reflects only the core energy, which can deviate from $E_{\rm ISO, w}$ inferred from modeling the late-time afterglow when the wing dominates. A similar mismatch can arise when comparing true energies at different epochs, with prompt-based estimates probing the core and late-time afterglow estimates reflecting the wing.

Can two-component jets produce observational signatures that uniquely distinguish them from other re-brightening scenarios, such as energy injection or radially stratified ejecta? This is a challenging question, given the large number of model parameters involved. A potential signature of two-component jets is a chromatic break during the transition from core to wing emission. If the two components occupy different spectral regimes or have different values of $p$, the emergence of the wing may be accompanied by an observable change in the spectral slope. After the shock-crossing phase, both components evolve in the same external medium, and their characteristic frequencies follow similar scalings, differing primarily in their energy normalization. Thus, chromatic breaks may indicate a jump in the energies of the core and wing. Variations in $\varepsilon_e$, $\varepsilon_B$, or $p$ between the components can further enhance this effect. A possible example is GRB 191221B, which has been interpreted as a two-component jet with differing microphysical parameters to account for the observed spectral evolution \citep{Chen24GRB191221B}. At the same time, energy injection also increases the jet energy, which can shift the characteristic synchrotron frequencies and induce spectral evolution. Robustly discriminating between these scenarios likely requires detailed, event-by-event modeling, such as MCMC analyses. Even then, degeneracies may persist. For example, the late-time re-brightening in GRB 250221A has been interpreted both in terms of a two-component jet \citep{Tian26Short} and late energy injection \citep{AnguloValdez26Evidence}. 

Finally, an additional potentially distinctive feature of two-component jets is the presence of double jet breaks. In principle, an on-axis observer could first detect the core's jet break, followed at later times by a second one associated with the wing. However, if the wing is initially beamed away ($\Gamma_{0,\rm w} > \theta_{\rm n}^{-1}$), it becomes visible around the time of the core jet break. This effectively masks the first jet break, leaving only a later break of the wing. In contrast, if the wing is slower, its contribution peaks at later times and may overtake the core only after the core jet break has already occurred. This allows the possibility of observing two distinct jet breaks.

It has also been suggested that such breaks could manifest differently across different bands. For instance, this interpretation has been proposed for GRB 230815A \cite{Leung2025}, where the early X-ray break is attributed to the core and a much later radio break to the wing. However, our results do not support a scenario in which the wing remains subdominant in X-rays while dominating in radio. The relative contribution of the wing is primarily governed by the jet dynamics rather than strong band-dependent effects. Once the wing flux becomes dominant, it does so approximately achromatically. Consequently, a wing capable of obscuring the core's break and producing a late-time break in the radio band would also be expected to contribute similarly in X-rays.

Our work has several limitations. The two-component jet model assumes a sharp transition between the jet core and the wing, which is likely an oversimplification; in reality, the jet is expected to possess a continuous angular structure. We further restrict our analysis to on-axis observers. In addition, we assume identical, time-independent microphysical parameters and a simplified external medium, and neglect possible hydrodynamic effects such as lateral spreading and component interaction. Addressing all these limitations simultaneously would substantially increase the scope of the paper; we therefore defer a more comprehensive treatment to future work.

\begin{acknowledgments}
We thank Yihan Wang for useful discussions. 
This research was funded by the Science Committee of the Ministry of Science and Higher Education of the Republic of Kazakhstan (Grant No. AP26103591). EA acknowledges support by the Nazarbayev University Faculty Development Competitive Research Grant Program (no. 040225FD4713). PB's work was funded by a NASA grant 80NSSC24K0770, a grant (no. 2020747) from the United States-Israel Binational Science Foundation (BSF), Jerusalem, Israel and by a grant (no. 1649/23) from the Israel Science Foundation.
AI-based language tools were used to improve the clarity and grammar of the manuscript.
\end{acknowledgments}

\appendix

\section{Analytical estimates}
\label{sec:formulas}

\begin{figure}
    \centering
    \includegraphics[width=\columnwidth]{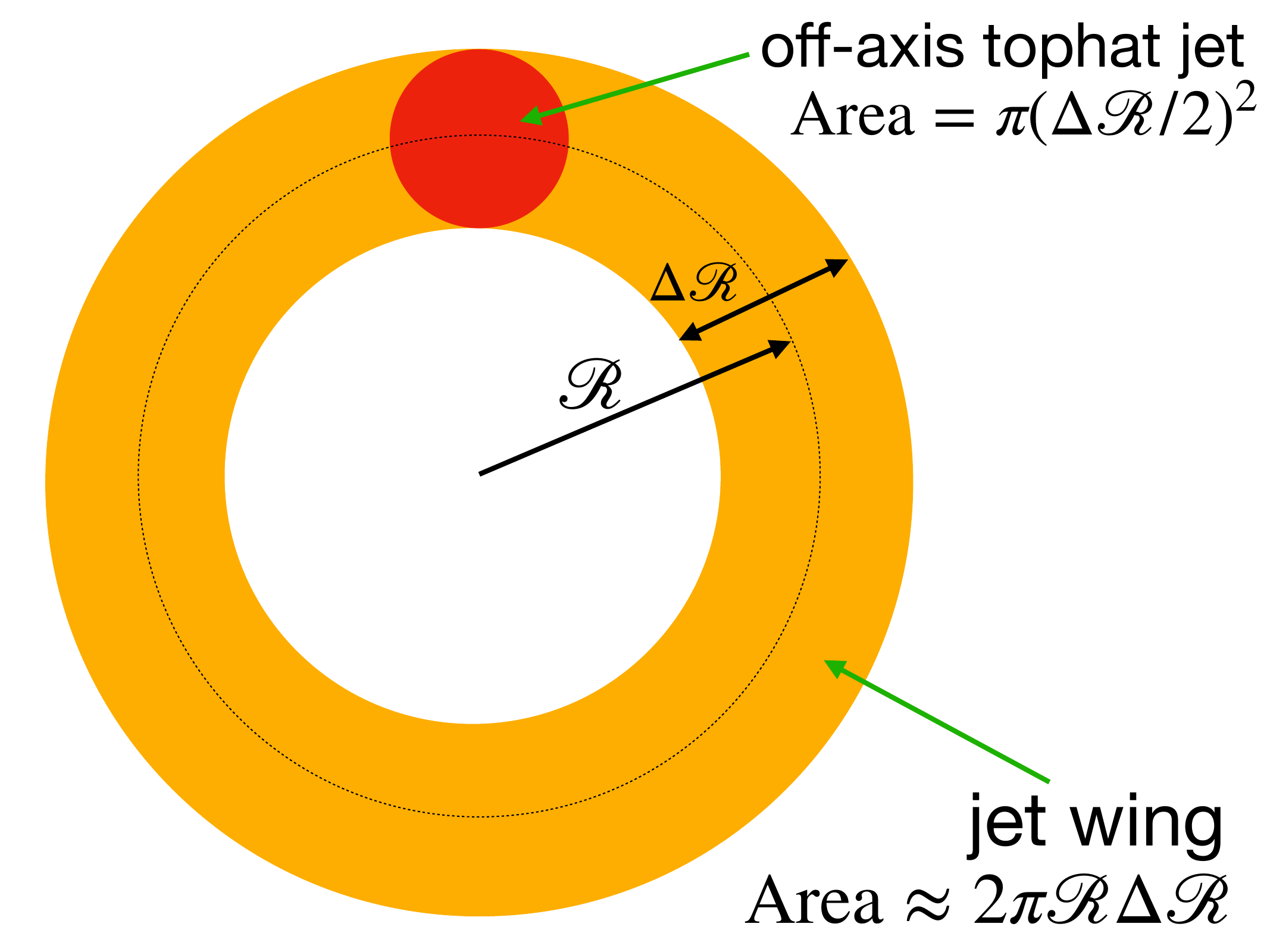}
    \caption{Schematic plot of off-axis jet model inscribed within the ring that represents the wing of two-component jet. The jet symmetry axis is perpendicular to the plane of this plot and passes through the center of the ring.}
    \label{fig:wing-jet-scheme}
\end{figure}

Since the analytical formalism describing emission from off-axis tophat jets is well developed \citep[e.g.,][]{Beniamini23Swift}, we approximate the emission from the jet wing as that from an off-axis tophat jet inscribed within the wing, whose opening angle equals the wing’s angular thickness. Because the emitting areas differ, the resulting flux must be rescaled accordingly. 

To estimate this factor, consider a flat ring with average radius $\mathcal{R}$ and thickness $\Delta\mathcal{R}$, representing an approximation of the wing’s front as viewed by an on-axis observer (Fig.~\ref{fig:wing-jet-scheme}). The area of ring is $2\pi \mathcal{R}\Delta\mathcal{R}$. Now consider a circle inscribed within the ring such that its diameter equals the ring thickness, as shown in Fig.~\ref{fig:wing-jet-scheme}. This circle approximates the front of the off-axis jet. Its area is $\pi(\Delta\mathcal{R}/2)^2$. The ratio of the ring area to the circle area is therefore $8\mathcal{R}/\Delta\mathcal{R}$.
In angular terms, $\mathcal{R} \propto (\theta_{\rm w}+\theta_{\rm n})/2$ and $\Delta\mathcal{R}\propto \theta_{\rm w}-\theta_{\rm n}$, yielding an area ratio of
\begin{equation}
    \mathcal{F} \approx 4\frac{\theta_w + \theta_n}{\theta_w - \theta_n}. \label{eq:F}
\end{equation}
We therefore multiply the flux from the off-axis tophat jet (whose opening angle equals the wing thickness) by this factor to account for the larger emitting area of the wing.

Another correction factor is required for the Doppler factor suppression in the off-axis regime. We apply \citep{Granot02Off}
\begin{align}
    a(\Gamma_{\rm w}) &\equiv \frac{\mathcal{D}(\theta_{\rm n})}{\mathcal{D}(0)} \approx \frac{1}{1+\Gamma_{\rm w}^2\theta_{\rm n}^2}, \label{eq:D} \\
    F_\nu^{\rm off}(t) &= a^3F_{\nu/a}^{\rm on}(at), \label{eq:Off}
\end{align}
where $\mathcal{D}(\theta_{\rm n})$ denotes the Doppler factor of the wing's inner edge viewed by the on-axis observer, and $\mathcal{D}(0)$ is the Doppler factor when viewed by a hypothetical observer aligned with the edge. We focus on the light curves of the optical R-band that resides in the regime $\text{max}(\nu_\mathrm{m}, \nu_\mathrm{a}) < \nu < \nu_\mathrm{c}$. In this case, we have
\begin{align}
    F_\nu &\propto \nu^{-\frac{p-1}{2}}, \\
    F_\nu^{\rm off}(t) &= a^{\frac{p+5}{2}}F_{\nu}^{\rm on}(at).
\end{align}
It's useful to derive the temporal slope of the flux before peak, as the wing's contribution can cause rebrightening. The rebrightening slope will be similar to the wing's flux slope. We have
\begin{align}
    \frac{d\log F_\nu^{\rm off}(t)}{d\log t} &= \frac{d\log F_{\nu}^{\rm on}(at)}{d\log (at)} - \frac{2\Gamma_{\rm w}^2\theta_{\rm n}^2}{1 + \Gamma_{\rm w}^2\theta_{\rm n}^2 \left(1 + 2\frac{d\log\Gamma_{\rm w}(at)}{d\log(at)}\right)} \nonumber \\
        &\times \frac{d\log\Gamma_{\rm w}(at)}{d\log(at)}\left( \frac{p+5}{2} + \frac{d\log F_{\nu}^{\rm on}(at)}{d\log (at)} \right),
\end{align}
where $\Gamma_{\rm w}(at)$ is treated as a function of time in the frame of an observer aligned with the wing's inner edge.

We use the following definition of the thickness parameter $\xi$:
\begin{align}
    l &= \left(\frac{3E}{4\pi nm_pc^2}\right)^{1/3}, \\
    \Delta_0 &= cT_{\rm eng}, \\
    \xi &\equiv \left(\frac{l}{\Delta_0}\right)^{1/2}\Gamma_0^{-4/3} = \left(\frac{3E}{4\pi nm_pc^5}\right)^{1/6} T_{\rm eng}^{-1/2} \Gamma_0^{-4/3},
\end{align}
where $T_{\rm eng}$ is the engine duration. Thin shells correspond to $\xi > 1$, in which the shell is thin compared to the deceleration length and the reverse shock remains Newtonian. Thick shells correspond to $\xi < 1$, in which the shell is thick and the reverse shock becomes relativistic \citep{Sari95AHydrodynamic, Kobayashi99Hydrodynamics}.

The four relevant length scales that described the shell evolution are defined as \citep{Sari95AHydrodynamic}:
\begin{itemize}
    \item $R_s$: the spreading radius, where the shell begins to expand significantly.
    \item $R_\Delta$: the shell-crossing radius, where the reverse shock finishes crossing the shell.
    \item $R_\Gamma$: the deceleration radius, where the bulk of the shell has swept up enough external medium to begin decelerating.
    \item $R_N$: the radius at which the reverse shock becomes relativistic.
\end{itemize}
These radii are related through the shell thickness parameter $\xi$:
\begin{equation}
    \frac{R_N}{\xi} = R_\Gamma = \sqrt{\xi}R_\Delta = \xi^2R_s. \label{eq:radii_order}
\end{equation}
The hierarchy of these radii directly determines the temporal evolution of the Lorentz factor $\Gamma$.

\subsection{ISM: Thin-Shell FS}

In the thin-shell regime, the ordering of the radii in Eq.~(\ref{eq:radii_order}) becomes
\begin{equation}
    R_s < R_\Delta < R_\Gamma < R_N,
\end{equation}
which means that the reverse shock crosses the shell before the deceleration begins \citep{Sari97Hydrodynamics}. Consequently, the Lorentz factor stays constant until $t_{\rm dec}$, after which it decays as follows:
\begin{equation}
    \Gamma \propto
    \begin{cases}
        \Gamma_0, & t < t_{\rm dec} \\
        E^{1/8} t^{-3/8}, & t > t_{\rm dec}
    \end{cases},
\end{equation}
where
\begin{equation}
    t_{\rm dec} \propto E^{1/3}_{\rm ISO} \Gamma_0^{-8/3}. \label{eq:tdec}
\end{equation}
The critical frequencies scale as
\begin{align}
    \nu_\mathrm{m} &\propto
    \begin{cases}
        \Gamma_0^4, & t < t_{\rm dec} \\
        E^{1/2}_{\rm ISO} t^{-3/2}, & t > t_{\rm dec}
    \end{cases}, \\
    \nu_\mathrm{c} &\propto
    \begin{cases}
        \Gamma_0^{-4} t^{-2}, & t < t_{\rm dec} \\
        E^{-1/2}_{\rm ISO} t^{-1/2}, & t > t_{\rm dec}
    \end{cases}.
\end{align}
For $\nu_\mathrm{a} < \nu_\mathrm{m}$,
\begin{equation}
    \nu_\mathrm{a} \propto
    \begin{cases}
        \Gamma_0^{8/5} t^{3/5}, & t < t_{\rm dec} \\
        E^{1/5}_{\rm ISO}, & t > t_{\rm dec}
    \end{cases},
\end{equation}
and for $\nu_\mathrm{m} < \nu_\mathrm{a}$,
\begin{equation}
    \nu_\mathrm{a}(t>t_{\rm dec}) \propto
    \begin{cases}
        \Gamma_0^{\frac{4(p+2)}{p+4}} t^{\frac{2}{p+4}}, & t < t_{\rm dec} \\
        E^{\frac{p+2}{2(p+4)}}_{\rm ISO} t^{-\frac{3p+2}{2(p+4)}}, & t > t_{\rm dec}
    \end{cases}.
\end{equation}
The spectral peak flux scales as
\begin{equation}
    F_{\nu,\rm{max}} \propto
    \begin{cases}
        \Gamma_0^8 t^3, & t < t_{\rm dec} \\
        E_{\rm ISO}, & t > t_{\rm dec}
    \end{cases}.
\end{equation}
The flux of a band in the regime $\text{max}(\nu_\mathrm{m},\nu_\mathrm{a}) < \nu < \nu_\mathrm{c}$ follows these scalings:
\begin{equation}
    F_\nu \propto \nu^{-\frac{p-1}{2}} \times
    \begin{cases}
         t^3, & t < t_{\rm dec} \\
        t^{-\frac{3(p-1)}{4}}, & t > t_{\rm dec}
    \end{cases}.
\end{equation}
For a thin-shell wing, the Doppler factor correction is expressed as
\begin{equation}
    a(\Gamma_{\rm w}) \Big|_{t_{\rm peak}} \approx
    \begin{cases}
        1/(1 + \Gamma_{0,\rm w}^2\theta_{\rm n}^2), & \Gamma_{0,\rm w} < 1/\theta_{\rm n} \\
        1/2, & \Gamma_{0,\rm w} > 1/\theta_{\rm n}
    \end{cases}.
\end{equation}
The wing-to-core peak time ratio in the observer's frame is given by
\begin{equation}
\begin{split}
    \mathcal{R}(t_{\rm peak}) &= \mathcal{R}(E_{\rm ISO})^{1/3} \times \\
    &\begin{cases}
        \mathcal{R}(\Gamma_0)^{-8/3} (1 + \Gamma_{0,\rm w}^2\theta_{\rm n}^2), & \Gamma_{0,\rm w} < 1/\theta_{\rm n} \\
       2(\Gamma_{0,\rm n}\theta_{\rm n})^{8/3}, & \Gamma_{0,\rm w} > 1/\theta_{\rm n}
    \end{cases}.
\end{split}
\end{equation}
The ratio of the corresponding peak fluxes becomes
\begin{equation}
    \mathcal{R}(F_{\nu,\rm peak}) = \mathcal{F}\mathcal{R}(E_{\rm ISO}) \times
    \begin{cases}
        a(\Gamma_{0,\rm w})^{\frac{p+5}{2}}\mathcal{R}(\Gamma_0)^{2(p-1)} \\
        a(1/\theta_{\rm n})^{\frac{p+5}{2}} (\Gamma_{0,\rm n}\theta_{\rm n})^{2(1-p)}
    \end{cases}.
\end{equation}
The slope of the wing's pre-peak flux can be expressed as
\begin{align}
    \frac{d\log F_\nu^{\rm w}}{d\log t} \Bigg|_{t < t_{\rm peak}} &=
        \begin{cases}
            3 \\
            -\frac{3(p-1)}{4} + \frac{3(13-p)}{4}\frac{\Gamma_{\rm w}^2\theta_{\rm n}^2}{4 + \Gamma_{\rm w}^2\theta_{\rm n}^2}
        \end{cases} \nonumber \\
        &\approx
        \begin{cases}
            3 & (t < t_{\rm dec, w}) \\
            -1 + 8\frac{\Gamma_{\rm w}^2\theta_{\rm n}^2}{4 + \Gamma_{\rm w}^2\theta_{\rm n}^2} & (t_{\rm dec} < t < t_{\rm beam, w})
        \end{cases}.
\end{align}
In the limit of $\Gamma_{\rm w}\theta_{\rm n}\gg1$, the slope approaches the value of $7$.

\subsection{ISM: Thin-Shell RS}

The blastwave dynamics is the same as in the thin-shell FS case during the shock-crossing phase. After that, the shocked ejecta gradually adjusts to the BM solution.

The analytical expressions detailed below represent the standard scaling laws for thin-shell RS emission, derived under the assumption that the shock reaches mildly relativistic speeds \citep{Sari95AHydrodynamic, Kobayashi00Light}. However, in the more precise numerical framework of VegasAfterglow, the reverse shock remains Newtonian throughout the entire shell \citep{Wang26VegasAfterglow}. Therefore, the injected electrons are deep-Newtonian ($\gamma_m\approx1$), which introduces corrections to the scalings commonly presented in the literature \citep{Sari95AHydrodynamic, Kobayashi00Light, Gao13complete}. In particular, the scaling for the injection frequency $\nu_m$ is affected since $\nu_m \propto \gamma_m^2$. Consequently, the scalings for the self-absorption frequency $\nu_a$ change as well. Additionally, since the electrons have low energy, a significant portion of them lies in the cyclotron regime, where synchrotron emission is less efficient \citep{Wang26VegasAfterglow}. Therefore, the shocked number of electrons is suppressed by the following factor:
\begin{equation}
    f_{\rm syn} = \left(\frac{\gamma_m-1}{\gamma_m}\right)^\kappa,
\end{equation}
where
\begin{equation}
    \kappa =
    \begin{cases}
        1, & 2<p<3 \\
        (p-1)/2, & p>3
    \end{cases}.
\end{equation}
This factor suppresses the emission and affects its temporal slope. These corrections contribute to the discrepancy between the numerical scalings and the standard analytical expressions presented below.

According to standard models, the critical frequencies scale as
\begin{align}
    \nu_\mathrm{m} &\propto
    \begin{cases}
        E^{-2}_{\rm ISO} \Gamma_0^{18} t^6, & t < t_{\rm dec} \\
        E^{18/35}_{\rm ISO} \Gamma_0^{-74/35} t^{-54/35}, & t > t_{\rm dec}
    \end{cases}, \\
    \nu_\mathrm{c} &\propto
    \begin{cases}
        \Gamma_0^{-4} t^{-2}, & t < t_{\rm dec} \\
        E^{-16/105}_{\rm ISO} \Gamma_0^{-292/105} t^{-54/35}, & t > t_{\rm dec}
    \end{cases}.
\end{align}
For $\nu_\mathrm{a} < \nu_\mathrm{m}$,
\begin{equation}
    \nu_\mathrm{a} \propto
    \begin{cases}
        E^{13/10}_{\rm ISO} \Gamma_0^{-36/5} t^{-33/10}, & t < t_{\rm dec} \\
        \Gamma_0^{8/175} t^{-102/175}, & t > t_{\rm dec}
    \end{cases},
\end{equation}
and for $\nu_\mathrm{m} < \nu_\mathrm{a}$,
\begin{equation}
    \nu_\mathrm{a} \propto
    \begin{cases}
        E^{\frac{3-2p}{p+4}}_{\rm ISO} \Gamma_0^{\frac{18p-12}{p+4}} t^{\frac{6p-7}{p+4}}, & t < t_{\rm dec} \\
        E^{\frac{2(9p+29)}{35(p+4)}}_{\rm ISO} \Gamma_0^{\frac{-74p-44}{35(p+4)}} t^{-\frac{54p+104}{35(p+4)}}, & t > t_{\rm dec}
    \end{cases}.
\end{equation}
The spectral peak flux scales as
\begin{equation}
    F_{\nu,\rm{max}} \propto
    \begin{cases}
        E^{1/2}_{\rm ISO} \Gamma_0^5 t^{3/2}, & t < t_{\rm dec} \\
        E^{139/105}_{\rm ISO} \Gamma_0^{-167/105} t^{-34/35}, & t > t_{\rm dec}
    \end{cases}.
\end{equation}
The flux of a band in the regime $\text{max}(\nu_\mathrm{m},\nu_\mathrm{a}) < \nu < \nu_\mathrm{c}$ follows these scalings:
\begin{equation}
    F_\nu \propto \nu^{-\frac{p-1}{2}} \times
    \begin{cases}
        t^{\frac{3(2p-1)}{2}}, & t < t_{\rm dec} \\
        t^{-\frac{27p+7}{35}}, & t > t_{\rm dec}
    \end{cases}.
\end{equation}
The peak time and flux ratios become
\begin{align}
    \mathcal{R}(t_{\rm peak}) &= \frac{1}{a(\Gamma_{0,\rm w})}\mathcal{R}(E_{\rm ISO})^{1/3} \mathcal{R}(\Gamma_0)^{-8/3}, \\
    \mathcal{R}(F_{\nu,\rm peak}) &= a(\Gamma_{0,\rm w})^\frac{p+5}{2}\mathcal{F}\mathcal{R}(E_{\rm ISO}) \mathcal{R}(\Gamma_0)^{p}.
\end{align}
Since the Lorentz factor stays constant during the shock-crossing phase in the thin-shell regime, the debeaming contribution to the slope is absent. The pre-peak slopes for the core and the wing are the same and given by
\begin{equation}
    \frac{d\log F_\nu}{d\log t} \Bigg|_{t < t_{\rm peak}} = \frac{3(2p-1)}{2} = 5.4.
\end{equation}

\subsection{ISM: Thick-Shell FS}

In the thick-shell case, the radii in Eq.~(\ref{eq:radii_order}) are ordered in reverse \citep{Sari95AHydrodynamic}:
\begin{equation}
    R_s > R_\Delta > R_\Gamma > R_N.
\end{equation}
The reverse shock becomes relativistic early, which causes the blastwave to decelerate during the shock-crossing phase. The deceleration radius $R_\Gamma$ is no longer relevant. The blastwave decelerates as
\begin{equation}
    \Gamma \propto
    \begin{cases}
        \Gamma_0\xi_0^{3/4}t^{-1/4}, & t < t_\times \\
        E^{1/8}_{\rm ISO} t^{-3/8}, & t > t_\times
    \end{cases},
\end{equation}
which at $t_\times$ becomes
\begin{equation}
    \Gamma(t_\times) \sim \left(\frac{3E_{\rm ISO}}{4\pi nm_pc^5}\right)^{1/8} T_{\rm eng}^{-3/8}.
\end{equation}
This value does not depend on the initial value $\Gamma_0$.

The peak flux scales as
\begin{equation}
    F_{\nu,\rm{max}} \propto E_{\rm ISO}\left(\frac{t}{t_\times}\right),
\end{equation}
In the thick-shell case, we have $T_{\rm eng} > t_{\rm dec}$. Therefore, $t_\times \approx T_{\rm eng}$ for the on-axis observer.

During the crossing phase, the critical frequencies scale as
\begin{align}
    \nu_\mathrm{m}(t<t_\times) &\propto E^{1/2}_{\rm ISO} t_\times^{-1/2} t^{-1}, \\
    \nu_\mathrm{c}(t<t_\times) &\propto E^{-1/2}_{\rm ISO} t_\times^{1/2} t^{-1}.
\end{align}
The absorption frequency:
\begin{equation}
    \nu_\mathrm{a}(t<t_\times) \propto
    \begin{cases}
        E^{1/5}_{\rm ISO} t_\times^{-1/5} t^{1/5}, & \nu_\mathrm{a} < \nu_\mathrm{m} \\
        E^{\frac{p+2}{2(p+4)}}_{\rm ISO} t_\times^{-\frac{p+2}{2(p+4)}} t^{-\frac{p}{p+4}}, & \nu_\mathrm{m} < \nu_\mathrm{a}
    \end{cases}.
\end{equation}
The flux scales as
\begin{equation}
    F_\nu \propto \nu^{-\frac{p-1}{2}} t^{\frac{3-p}{2}},
\end{equation}
from which the pre-peak slope can be identified as $\frac{3-p}{2}$.
The peak flux is proportional to
\begin{equation}
    F^{\rm n}_{\nu,\rm peak} \propto E_{\rm ISO,n}^{\frac{p+3}{4}} T_{\rm eng}^{-\frac{3(p-1)}{4}}.
\end{equation}
If the wing is still beamed away from the observer at the shock-crossing time, its peak time and flux occur at the debeaming time $t_{\rm beam, w}$. The expressions for these quantities can be written as
\begin{align}
    t_{\rm peak,w} &\propto a(1/\theta_{\rm n})^{-1}E_{\rm ISO,w}^{1/3} \theta_{\rm n}^{8/3}, \\
    F^{\rm w}_{\nu,\rm peak} &\propto a(1/\theta_{\rm n})^{\frac{p+5}{2}}\mathcal{F}E_{\rm ISO,w} \theta_{\rm n}^{2(1-p)}.
\end{align}
The wing's pre-peak slope becomes
\begin{equation}
    \frac{d\log F_\nu^{\rm w}}{d\log t} \Bigg|_{t < t_{\times,\rm w}} = \frac{3-p}{2} + \frac{4\Gamma_{\rm w}^2\theta_{\rm n}^2}{2 + \Gamma_{\rm w}^2\theta_{\rm n}^2} < 4.3,
\end{equation}
where $4.3$ is reached in the limit $\Gamma_{\rm w}\theta_{\rm n}\gg1$.

\subsection{ISM: Thick-Shell RS}

In this case, the RS becomes relativistic early. The blastwave decelerates as
\begin{equation}
    \gamma_3 \sim \frac{\Gamma_0\xi_0^{3/4}}{\sqrt2} \times
    \begin{cases}
        (t / t_\times)^{-1/4}, & t < t_\times \\
        (t / t_\times)^{-7/16}, & t > t_\times
    \end{cases},
\end{equation}
The scalings laws for the thick-shell RS are given as
\begin{align}
    F_{\nu,\rm{max}} &\propto E^{5/4}_{\rm ISO}\Gamma_0^{-1} \times
    \begin{cases}
        t_\times^{-5/4}  t^{1/2}, & t < t_\times \\
        t_\times^{11/48} t^{-47/48}, & t > t_\times
    \end{cases}, \\
    \nu_\mathrm{m} &\propto \Gamma_0^2 \times
    \begin{cases}
        1, & t < t_\times \\
        t_\times^{73/48} t^{-73/48}, & t > t_\times
    \end{cases}, \\
    \nu_\mathrm{c} &\propto E^{-1/2}_{\rm ISO} \times
    \begin{cases}
        t_\times^{1/2} t^{-1}, & t < t_\times \\
        t_\times^{49/48} t^{-73/48}, & t > t_\times
    \end{cases}.
\end{align}
For $\nu_\mathrm{a} < \nu_\mathrm{m}$,
\begin{equation}
    \nu_\mathrm{a} \propto E^{3/5}_{\rm ISO}\Gamma_0^{-8/5} \times
    \begin{cases}
        t_\times^{-3/5} t^{-3/5}, & t < T_{\rm eng} \\
        t_\times^{-2/3} t^{-8/15}, & t > T_{\rm eng}
    \end{cases},
\end{equation}
and for $\nu_\mathrm{m} < \nu_\mathrm{a}$,
\begin{equation}
    \nu_\mathrm{a} \propto E^{\frac{2}{p+4}}_{\rm ISO}\Gamma_0^{\frac{2(p-2)}{p+4}} \times
    \begin{cases}
        t_\times^{-\frac{p}{p+4}} t^{-\frac{2}{p+4}} \\
        t_\times^{\frac{73p-58}{48(p+4)}} t^{-\frac{73p+134}{48(p+4)}}
    \end{cases}.
\end{equation}
The flux scales as
\begin{equation}
    F_{\nu,\rm max} \propto \nu^{-\frac{p-1}{2}}
    \begin{cases}
        t^{1/2}, & t < T_{\rm eng} \\
        t^{-\frac{73p+21}{96}}, & t > T_{\rm eng}
    \end{cases}.
\end{equation}
The core's peak occurs at $T_{\rm eng}$, and its peak flux is proportional to
\begin{equation}
    F^{\rm n}_{\nu,\rm peak} \propto E_{\rm ISO, n}^{5/4} \Gamma_{0,\rm n}^{p-2} T_{\rm eng}^{-3/4}.
\end{equation}
The wing's emission is delayed in the observer's frame. Therefore, its peak time and flux are given by
\begin{align}
    t_{\times,\rm w} &= T_{\rm eng} \times a(\Gamma_{\rm w}(t_{\times,\rm w}))^{-1}, \\
    F_{\nu, \rm peak} &\propto a(\Gamma_{\rm w}(t_{\times,\rm w}))^{\frac{p+5}{2}} \mathcal{F} E_{\rm ISO, w}^{5/4} \Gamma_{0,\rm w}^{p-2} T_{\rm eng}^{-3/4}.
\end{align}
Since the Lorentz factor decelerates during the shock-crossing time in the thick-shell regime, the debeaming effect enhances the wing's pre-peak slope as
\begin{equation}
    \frac{d\log F_\nu^{\rm w}}{d\log t} \Bigg|_{t < t_{\rm peak}} = \frac{1}{2} + \frac{\Gamma_{\rm w}^2\theta_{\rm n}^2}{2 + \Gamma_{\rm w}^2\theta_{\rm n}^2} \left(\frac{p}{2} + 3\right) < 4.6,
\end{equation}
approaching $4.6$ in the limit $\Gamma_{\rm w}\theta_{\rm n}\gg1$.

\subsection{Wind}

\begin{figure}[htbp]
    \centering

    \begin{subfigure}{\columnwidth}
        \centering
        \begin{tikzpicture}[thick, scale=0.6]
            \draw[->] (0,0) -- (10,0) node[right] {$\nu$};

            \node[left] at (-0.5, 0.8) {\footnotesize $t<t_\times$};
            \node[left] at (-0.5, -0.8) {\footnotesize $t>t_\times$};

            \foreach \x/\l in {2.5/\nu_\mathrm{a}, 5/\nu_\mathrm{c}, 7.5/\nu_\mathrm{m}} {
                \draw (\x,0.1) -- (\x,-0.1);
                \node[below=5pt] at (\x,0) {$\l$};
            }

            \node at (1.25, 0.8) {\footnotesize $3$};
            \node at (3.75, 0.8) {\footnotesize $-\frac{1}{3}$};
            \node at (6.25, 0.8) {\footnotesize $\frac{1}{2}$};
            \node at (8.75, 0.8) {\footnotesize $-\frac{p-2}{2}$};

            \node at (1.25, -0.8) {\footnotesize $2$};
            \node at (3.75, -0.8) {\footnotesize $-\frac{2}{3}$};
            \node at (6.25, -0.8) {\footnotesize $-\frac{1}{4}$};
            \node at (8.75, -0.8) {\footnotesize $-\frac{3p-2}{4}$};
        \end{tikzpicture}
    \end{subfigure}

    \vspace{0.4cm}

    \begin{subfigure}{\columnwidth}
        \centering
        \begin{tikzpicture}[thick, scale=0.6]
            \draw[->] (0,0) -- (10,0) node[right] {$\nu$};

            \node[left] at (-0.5, 0.8) {\footnotesize $t<t_\times$};
            \node[left] at (-0.5, -0.8) {\footnotesize $t>t_\times$};

            \foreach \x/\l in {2.5/\nu_\mathrm{a}, 5/\nu_\mathrm{m}, 7.5/\nu_\mathrm{c}} {
                \draw (\x,0.1) -- (\x,-0.1);
                \node[below=5pt] at (\x,0) {$\l$};
            }

            \node at (1.25, 0.8) {\footnotesize $2$};
            \node at (3.75, 0.8) {\footnotesize $\frac{1}{3}$};
            \node at (6.25, 0.8) {\footnotesize $-\frac{p-1}{2}$};
            \node at (8.75, 0.8) {\footnotesize $-\frac{p-2}{2}$};

            \node at (1.25, -0.8) {\footnotesize $1$};
            \node at (3.75, -0.8) {\footnotesize $0$};
            \node at (6.25, -0.8) {\footnotesize $-\frac{3p-1}{4}$};
            \node at (8.75, -0.8) {\footnotesize $-\frac{3p-2}{4}$};
        \end{tikzpicture}
    \end{subfigure}

    \vspace{0.4cm}

    \begin{subfigure}{\columnwidth}
        \centering
        \begin{tikzpicture}[thick, scale=0.6]
            \draw[->] (0,0) -- (10,0) node[right] {$\nu$};

            \node[left] at (-0.5, 0.8) {\footnotesize $t<t_\times$};
            \node[left] at (-0.5, -0.8) {\footnotesize $t>t_\times$};

            \foreach \x/\l in {2.5/\nu_\mathrm{m}, 5/\nu_\mathrm{a}, 7.5/\nu_\mathrm{c}} {
                \draw (\x,0.1) -- (\x,-0.1);
                \node[below=5pt] at (\x,0) {$\l$};
            }

            \node at (1.25, 0.8) {\footnotesize $2$};
            \node at (3.75, 0.8) {\footnotesize $\frac{5}{2}$};
            \node at (6.25, 0.8) {\footnotesize $-\frac{p-1}{2}$};
            \node at (8.75, 0.8) {\footnotesize $-\frac{p-2}{2}$};

            \node at (1.25, -0.8) {\footnotesize $1$};
            \node at (3.75, -0.8) {\footnotesize $\frac{7}{4}$};
            \node at (6.25, -0.8) {\footnotesize $-\frac{3p-1}{4}$};
            \node at (8.75, -0.8) {\footnotesize $-\frac{3p-2}{4}$};
        \end{tikzpicture}
    \end{subfigure}

    \caption{Temporal powers of FS flux for the wind medium.}
    \label{fig:Wind_FS_tikz}
\end{figure}
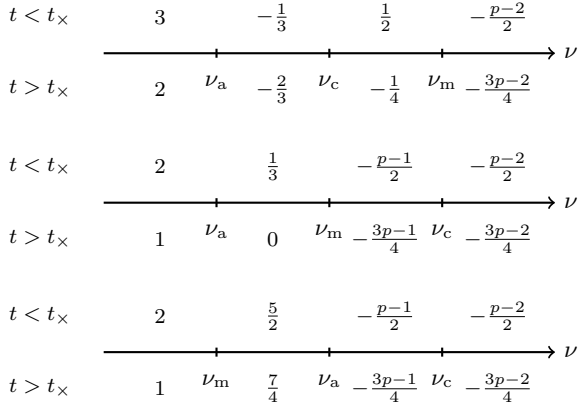

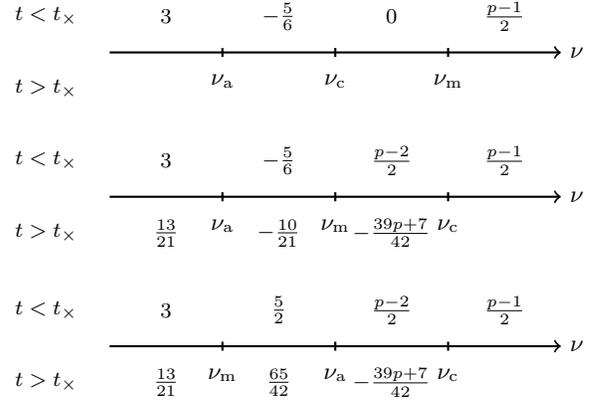
\begin{figure}[htbp]
    \centering

    \begin{subfigure}{\columnwidth}
        \centering
        \begin{tikzpicture}[thick, scale=0.6]
            \draw[->] (0,0) -- (10,0) node[right] {$\nu$};

            \node[left] at (-0.5, 0.8) {\footnotesize $t<t_\times$};
            \node[left] at (-0.5, -0.8) {\footnotesize $t>t_\times$};

            \foreach \x/\l in {2.5/\nu_\mathrm{a}, 5/\nu_\mathrm{c}, 7.5/\nu_\mathrm{m}} {
                \draw (\x,0.1) -- (\x,-0.1);
                \node[below=5pt] at (\x,0) {$\l$};
            }

            \node at (1.25, 0.8) {\footnotesize $3$};
            \node at (3.75, 0.8) {\footnotesize $-\frac{5}{6}$};
            \node at (6.25, 0.8) {\footnotesize $0$};
            \node at (8.75, 0.8) {\footnotesize $\frac{p-1}{2}$};

            \node at (1.25, -0.8) {\footnotesize };
            \node at (3.75, -0.8) {\footnotesize };
            \node at (6.25, -0.8) {\footnotesize };
            \node at (8.75, -0.8) {\footnotesize };
        \end{tikzpicture}
    \end{subfigure}

    \vspace{0.4cm}

    \begin{subfigure}{\columnwidth}
        \centering
        \begin{tikzpicture}[thick, scale=0.6]
            \draw[->] (0,0) -- (10,0) node[right] {$\nu$};

            \node[left] at (-0.5, 0.8) {\footnotesize $t<t_\times$};
            \node[left] at (-0.5, -0.8) {\footnotesize $t>t_\times$};

            \foreach \x/\l in {2.5/\nu_\mathrm{a}, 5/\nu_\mathrm{m}, 7.5/\nu_\mathrm{c}} {
                \draw (\x,0.1) -- (\x,-0.1);
                \node[below=5pt] at (\x,0) {$\l$};
            }

            \node at (1.25, 0.8) {\footnotesize $3$};
            \node at (3.75, 0.8) {\footnotesize $-\frac{5}{6}$};
            \node at (6.25, 0.8) {\footnotesize $\frac{p-2}{2}$};
            \node at (8.75, 0.8) {\footnotesize $\frac{p-1}{2}$};

            \node at (1.25, -0.8) {\footnotesize $\frac{13}{21}$};
            \node at (3.75, -0.8) {\footnotesize $-\frac{10}{21}$};
            \node at (6.25, -0.8) {\footnotesize $-\frac{39p+7}{42}$};
            \node at (8.75, -0.8) {\footnotesize };
        \end{tikzpicture}
    \end{subfigure}

    \vspace{0.4cm}

    \begin{subfigure}{\columnwidth}
        \centering
        \begin{tikzpicture}[thick, scale=0.6]
            \draw[->] (0,0) -- (10,0) node[right] {$\nu$};

            \node[left] at (-0.5, 0.8) {\footnotesize $t<t_\times$};
            \node[left] at (-0.5, -0.8) {\footnotesize $t>t_\times$};

            \foreach \x/\l in {2.5/\nu_\mathrm{m}, 5/\nu_\mathrm{a}, 7.5/\nu_\mathrm{c}} {
                \draw (\x,0.1) -- (\x,-0.1);
                \node[below=5pt] at (\x,0) {$\l$};
            }

            \node at (1.25, 0.8) {\footnotesize $3$};
            \node at (3.75, 0.8) {\footnotesize $\frac{5}{2}$};
            \node at (6.25, 0.8) {\footnotesize $\frac{p-2}{2}$};
            \node at (8.75, 0.8) {\footnotesize $\frac{p-1}{2}$};

            \node at (1.25, -0.8) {\footnotesize $\frac{13}{21}$};
            \node at (3.75, -0.8) {\footnotesize $\frac{65}{42}$};
            \node at (6.25, -0.8) {\footnotesize $-\frac{39p+7}{42}$};
            \node at (8.75, -0.8) {\footnotesize };
        \end{tikzpicture}
    \end{subfigure}

    \caption{Temporal powers of thin-shell RS flux for the wind medium.}
    \label{fig:Wind_RS_thin_tikz}
\end{figure}

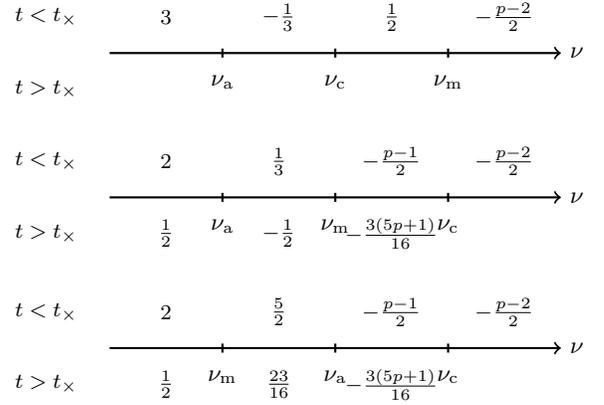
\begin{figure}[htbp]
    \centering

    \begin{subfigure}{\columnwidth}
        \centering
        \begin{tikzpicture}[thick, scale=0.6]
            \draw[->] (0,0) -- (10,0) node[right] {$\nu$};

            \node[left] at (-0.5, 0.8) {\footnotesize $t<t_\times$};
            \node[left] at (-0.5, -0.8) {\footnotesize $t>t_\times$};

            \foreach \x/\l in {2.5/\nu_\mathrm{a}, 5/\nu_\mathrm{c}, 7.5/\nu_\mathrm{m}} {
                \draw (\x,0.1) -- (\x,-0.1);
                \node[below=5pt] at (\x,0) {$\l$};
            }

            \node at (1.25, 0.8) {\footnotesize $3$};
            \node at (3.75, 0.8) {\footnotesize $-\frac{1}{3}$};
            \node at (6.25, 0.8) {\footnotesize $\frac{1}{2}$};
            \node at (8.75, 0.8) {\footnotesize $-\frac{p-2}{2}$};

            \node at (1.25, -0.8) {\footnotesize };
            \node at (3.75, -0.8) {\footnotesize };
            \node at (6.25, -0.8) {\footnotesize };
            \node at (8.75, -0.8) {\footnotesize };
        \end{tikzpicture}
    \end{subfigure}

    \vspace{0.4cm}

    \begin{subfigure}{\columnwidth}
        \centering
        \begin{tikzpicture}[thick, scale=0.6]
            \draw[->] (0,0) -- (10,0) node[right] {$\nu$};

            \node[left] at (-0.5, 0.8) {\footnotesize $t<t_\times$};
            \node[left] at (-0.5, -0.8) {\footnotesize $t>t_\times$};

            \foreach \x/\l in {2.5/\nu_\mathrm{a}, 5/\nu_\mathrm{m}, 7.5/\nu_\mathrm{c}} {
                \draw (\x,0.1) -- (\x,-0.1);
                \node[below=5pt] at (\x,0) {$\l$};
            }

            \node at (1.25, 0.8) {\footnotesize $2$};
            \node at (3.75, 0.8) {\footnotesize $\frac{1}{3}$};
            \node at (6.25, 0.8) {\footnotesize $-\frac{p-1}{2}$};
            \node at (8.75, 0.8) {\footnotesize $-\frac{p-2}{2}$};

            \node at (1.25, -0.8) {\footnotesize $\frac{1}{2}$};
            \node at (3.75, -0.8) {\footnotesize $-\frac{1}{2}$};
            \node at (6.25, -0.8) {\footnotesize $-\frac{3(5p+1)}{16}$};
            \node at (8.75, -0.8) {\footnotesize };
        \end{tikzpicture}
    \end{subfigure}

    \vspace{0.4cm}

    \begin{subfigure}{\columnwidth}
        \centering
        \begin{tikzpicture}[thick, scale=0.6]
            \draw[->] (0,0) -- (10,0) node[right] {$\nu$};

            \node[left] at (-0.5, 0.8) {\footnotesize $t<t_\times$};
            \node[left] at (-0.5, -0.8) {\footnotesize $t>t_\times$};

            \foreach \x/\l in {2.5/\nu_\mathrm{m}, 5/\nu_\mathrm{a}, 7.5/\nu_\mathrm{c}} {
                \draw (\x,0.1) -- (\x,-0.1);
                \node[below=5pt] at (\x,0) {$\l$};
            }

            \node at (1.25, 0.8) {\footnotesize $2$};
            \node at (3.75, 0.8) {\footnotesize $\frac{5}{2}$};
            \node at (6.25, 0.8) {\footnotesize $-\frac{p-1}{2}$};
            \node at (8.75, 0.8) {\footnotesize $-\frac{p-2}{2}$};

            \node at (1.25, -0.8) {\footnotesize $\frac{1}{2}$};
            \node at (3.75, -0.8) {\footnotesize $\frac{23}{16}$};
            \node at (6.25, -0.8) {\footnotesize $-\frac{3(5p+1)}{16}$};
            \node at (8.75, -0.8) {\footnotesize };
        \end{tikzpicture}
    \end{subfigure}

    \caption{Temporal powers of thick-shell RS flux for the wind medium.}
    \label{fig:Wind_RS_thick_tikz}
\end{figure}

The derivations in a wind-like medium follows the same procedure as in the ISM case: one combines the dynamical evolution of the shocked regions with the time dependence of the critical frequencies to obtain the flux scalings in a particular spectram segment of interest. Since the algebra is analogous, we do not repeat it here. Instead, we summarize the results in the diagrams presented in Figs.~\ref{fig:Wind_FS_tikz}-\ref{fig:Wind_RS_thick_tikz}. The diagrams present the temporal slopes of the FS and RS emission in a wind environment for all spectral regimes. In each diagram, we distinguish between the phases before and after the shock-crossing time, and list the corresponding temporal slopes in each segment.

The slopes shown here do not include the debeaming slope enhancement. In a wind medium, the Lorentz factor remains constant prior to shock crossing for both thin and thick shells (see Fig.~\ref{fig:Gamma_dur}), so the pre-crossing slopes are unaffected by debeaming. After shock crossing, the Lorentz factor begins to decrease, but more gradually than in the ISM case. As a result, the additional slope contribution from debeaming is comparatively weak; numerically, even for highly relativistic cases, it does not exceed an extra factor of $t^1$.

For the forward shock, the slopes for thin and thick shells are identical. We therefore combine both regimes into a single diagram (Fig.~\ref{fig:Wind_FS_tikz}). For the reverse shock, the behavior differs between thin and thick shells, and we present separate diagrams for each case in Fig.~\ref{fig:Wind_RS_thin_tikz} and~\ref{fig:Wind_RS_thick_tikz}, respectively.

\nocite{*}

\FloatBarrier
\providecommand{\noopsort}[1]{}\providecommand{\singleletter}[1]{#1}%

\end{document}